\documentclass[12pt]{article}

\usepackage[top=1in,left=1in,right=1in,bottom=1in]{geometry}

\usepackage[style=nature]{biblatex}
\usepackage{lscape}
\usepackage{setspace}

\usepackage{graphicx}
\usepackage{adjustbox}
\usepackage{booktabs}
\usepackage{longtable}
\usepackage{tabularx}
\usepackage[skip=0pt, labelfont=bf, font=footnotesize]{caption}
\usepackage{threeparttable}
\usepackage{threeparttablex}
\usepackage[capposition=top]{floatrow}
\newcounter{magicrownumbers}

\usepackage{nameref}
\usepackage{tikz}

\definecolor{c_color}{rgb}{0, 0, 0}
\definecolor{d_color}{rgb}{0, 0, 0}

\usepackage{times}

\begin{document} 

\pagebreak

\centerline{\LARGE The decline of disruptive science and technology}

\vspace{1cm}

\centerline{Michael Park,$^{1}$  Erin Leahey$^{2}$ Russell J. Funk,$^{1\ast}$}

\vspace{16cm}

\noindent\rule{14cm}{0.4pt} \\
\footnotesize{$^{1}$Carlson School of Management, University of Minnesota,}\\
\hspace*{2mm}\footnotesize{321 19th Ave S, Minneapolis, MN 55455, USA}\\
\footnotesize{$^{2}$Department of Sociology, University of Arizona}\\
\hspace*{2mm}\footnotesize{1145 E South Campus Drive, Tucson, AZ 85721, USA}\\
\vspace*{3mm}
\footnotesize{$^\ast$To whom correspondence should be addressed; e-mail:  rfunk@umn.edu.}

\pagebreak
\doublespacing

\subparagraph{Summary} \textbf{Theories of scientific and technological change view discovery and invention as endogenous  processes \autocite{fleming2001recombinant,schumpeter1942capitalism}, wherein prior accumulated knowledge enables future progress by allowing researchers to, in Newton’s words, “stand on the shoulders of giants”\autocite{koyre1952unpublished,weitzman1998recombinant,acemoglu2016innovation,popper2014conjectures,fleck2012genesis}. Recent decades have witnessed exponential growth in the volume of new scientific and technological knowledge, thereby creating conditions that should be ripe for major advances \autocite{tria2014dynamics,fink2017serendipity}. Yet contrary to this view, studies suggest that progress is slowing in several major fields of science and technology \autocite{pammolli2011productivity,bloom2020ideas}. Here, we analyze these claims at scale across 6 decades, using data on 45 million papers and 3.5 million patents from 6 large-scale datasets. We find that papers and patents are increasingly less likely to break with the past in ways that push science and technology in new directions, a pattern that holds universally across fields. Subsequently, we link this decline in disruptiveness to a narrowing in the use of prior knowledge, allowing us to reconcile the patterns we observe with the “shoulders of giants” view. We find that the observed declines are unlikely to be driven by changes in the quality of published science, citation practices, or field-specific factors. Overall, our results suggest that slowing rates of disruption may reflect a fundamental shift in the nature of science and technology.}

\vspace{3mm}

While the past century witnessed an unprecedented expansion of scientific and technological knowledge, there are concerns that innovative activity is slowing \autocite{jones2009burden,gordon2016rise,chu2021}. Studies document declining research productivity in semiconductors, pharmaceuticals, and other fields \autocite{pammolli2011productivity,bloom2020ideas}. Papers, patents, and even grant applications have become less novel and less likely to connect disparate areas of knowledge, both of which are precursors of innovation \autocite{packalen2020nih, jaffe2011innovation}. The gap between the year of discovery and the awarding of a Nobel Prize has also increased \autocite{horgan2015end,collison2018science}, suggesting that today’s contributions do not measure up to the past. These trends have attracted increasing attention from policymakers, as they pose significant threats to economic growth, human health and well being, and national security, along with global efforts to combat grand challenges like climate change\autocite{OECD2021effective,nola2021artificial}.

Numerous explanations for this slowdown have been proposed. Some point to a dearth of ``low hanging fruit'' as the readily available productivity-enhancing innovations have already been made \autocite{cowen2011great,gordon2016rise}. Others emphasize the increasing burden of knowledge; scientists and inventors require ever more training to reach the frontiers of their fields, leaving less time to push those frontiers forward\autocite{jones2009burden,einstein1949the}. Yet much remains unknown, not merely about the causes of slowing innovative activity, but also the depth and breadth of the phenomenon. The decline is difficult to reconcile with centuries of observation by philosophers of science, who characterize the growth of knowledge as an endogenous process, wherein prior knowledge enables future discovery, a view captured famously in Newton's observation that if he had seen further, it was by “standing on the shoulders of giants.”\autocite{koyre1952unpublished} Moreover, to date, the evidence pointing to a slowdown is based on studies of particular fields, using disparate and domain-specific metrics\autocite{pammolli2011productivity,bloom2020ideas}, making it difficult to know whether the changes are happening at similar rates across areas of science and technology. Little is also known about whether the patterns seen in aggregate indicators may mask differences in the degree to which individual works push the frontier. 

We address these gaps in understanding by analyzing 25 million papers (1945-2010) in the Web of Science (``WoS data'') and 3.5 million patents from (1976-2010) in the United States Patent and Trademark Office's Patents View database (``USPTO data''). The WoS data include 159 million citations and 28 million paper titles and abstracts. The USPTO data include 18 million citations and 6 million patent titles and abstracts. Subsequently, we replicate our core findings on 4 additional datasets---JSTOR, the American Physical Society corpus, Microsoft Academic Graph, and PubMed---encompassing 20 million papers. Using these data, we join a novel citation-based measure \autocite{funk2017dynamic} with textual analyses of titles and abstracts to understand whether papers and patents forge new directions over time and across fields.

To characterize the nature of innovation, we draw on foundational theories of scientific and technological change \autocite{schumpeter1942capitalism,anderson1990technological,arthur2007structure,}, which distinguish between two types of breakthroughs. First, some contributions improve existing streams of knowledge, and therefore consolidate the status quo. Kohn \& Sham (1965) \autocite{kohn1965self}, a Nobel-winning paper (``KS'') used established theorems to develop a method for calculating the structure of electrons, which cemented the value of prior research. Second, some contributions disrupt existing knowledge, rendering it obsolete, and propelling science and technology in new directions. Watson \& Crick (1953) \autocite{watson1953molecular} (``WC''), also a Nobel winner, introduced a model of the structure of DNA that superseded previous approaches (e.g., Pauling’s triple helix). KS and WC were both important, but their implications for scientific and technological change were different. 

We quantify this distinction using a measure---the CD index---that characterizes the consolidating or disruptive nature of science and technology (Figure 1). The intuition is that if a paper or patent is disruptive, the subsequent work that cites it is less likely to also cite its predecessors; for future researchers, the ideas that went into its production are less relevant (e.g., Pauling’s triple helix). If a paper or patent is consolidating, subsequent work that cites it is also more likely to cite its predecessors; for future researchers, the knowledge upon which the work builds is still (and perhaps more) relevant (e.g., the theorems KS used). The CD index ranges from -1 (consolidating) to 1 (disruptive). We measure the CD index five years after the year of each paper's publication (indicated by $CD_5$, see Extended Data Figure 1 for the distribution of $CD_5$ among papers and patents and Extended Data Figure 2 for analyses using alternative windows\autocite{bornmann2019length}). For example, WC and KS both received over a hundred citations within five years of being published. However, the KS paper has a $CD_5$ of -0.22 (indicating consolidation), whereas the WC paper has a $CD_5$ of 0.62 (indicating disruption). The CD index has been validated extensively in prior research, including through correlation with expert assessments \autocite{funk2017dynamic,wu2019large}.


Across fields, we find that science and technology are becoming less disruptive. Figure 2 plots the average $CD_5$ over time for papers (Figure 2A) and patents (Figure 2B). For papers, the decrease between 1945 and 2010 ranges from 91.9\% (where the average $CD_5$ dropped from 0.52 in 1945 to 0.04 in 2010 for the Social Sciences) to 100\% (where the average $CD_5$ decreased from 0.36 in 1945 to 0 in 2010 for the Physical Sciences); for patents, the decrease between 1980 and 2010 ranges from 78.7\% (where the average $CD_5$ decreased from 0.30 in 1980 to 0.06 in 2010 for Computers and Communications) to 91.5\% (where the average $CD_5$ decreased from 0.38 in 1980 to 0.03 in 2010 Drugs and Medical). For both papers and patents, the rates of decline are greatest in the earlier parts of the time series, and for patents, they appear to begin stabilizing between the years 2000 and 2005. For papers, since about 1980, the rate of decline has been more modest in the Life Sciences and Biomedicine and the Physical Sciences, and most dramatic and persistent in the Social Sciences and Technology. Overall, however, relative to earlier eras, recent papers and patents do less to push science and technology in new directions. The general similarity in trends we observe across fields is noteworthy in light of ``low hanging fruit'' theories \autocite{cowen2011great,gordon2016rise}, which would likely predict greater heterogeneity in the decline, as it seems unlikely fields would ``consume'' their low hanging fruit at similar rates/times.

The decline in disruptive science and technology is also observable using alternative indicators. Because they create departures from the status quo, disruptive papers and patents are likely to introduce new words (e.g., words used to create a new paradigm might differ from those that are used to develop an existing paradigm \autocite{kuhn1962structure,wray2011kuhn}). Therefore, if disruptiveness is declining, we would expect a decline in the diversity of words used in science and technology. To evaluate this, Figures 3A and 3D document the type-token ratio (i.e., unique/total words) of paper and patent titles over time (see Supplementary Information Section 1). We observe substantial declines, especially in the earlier periods, prior to 1970 for papers and 1990 for patents. For paper titles (Figure 3A), the decrease (1945-2010) ranges from 76.5\% (Social Science) to 88\% (Technology); for patent titles (Figure 3D), the decrease (1980-2010) ranges from 32.5\% (Chemical) to 81\% (Computer and Communications). For paper abstracts (Extended Data Figure 3A), the decrease (1992-2010) ranges from 23.1\% (Life Science and Biomedicine) to 38.9\% (Social Science); for patent abstracts (Extended Data Figure 3B), the decrease (1980-2010) ranges from 21.5\% (Mechanical) to 73.2\% (Computers and Communications). In Figures 3B and 3E, we demonstrate that these declines in word diversity are accompanied by similar declines in combinatorial novelty; over time, the particular words that scientists and inventors use in the titles of their papers and patents are increasingly likely to have been used together in the titles of prior work. Consistent with these trends in language, we also observe declining novelty in the combinations of prior work cited by papers and patents, based on a previously established measure of ``atypical combinations''\autocite{uzzi2013atypical} (Extended Data Figure 4).

The decline in disruptive activity is also apparent in the specific words used by scientists and inventors. If disruptiveness is declining, we reasoned that verbs alluding to the creation, discovery, or perception of new things should be used less frequently over time, whereas verbs alluding to the improvement, application, or assessment of existing things  may be used more often \autocite{kuhn1962structure,wray2011kuhn}. Figure 3 shows the most common verbs in paper (Figure 3C) and patent titles (Figure 3F) in the first and last decade of each sample (see also Supplementary Information Section 2). While precisely and quantitatively characterizing words as ``considating'' or ``disruptive'' is challenging in the absence of context, the figure highlights a clear and qualitative shift in language. In the earlier decades, verbs evoking creation (e.g., ``produce'', ``form'', ``prepare'', ``make''), discovery (e.g., ``determine'', ``report''), and perception (e.g., ``measure'') are prevalent in both paper and patent titles. In the later decades, however, these verbs are almost completely displaced by those tending to be more evocative of the improvement (e.g., ``improve'', ``enhance'', ``increase''), application (e.g., ``use'', ``include''), or assessment (e.g., ``associate'', ``mediate'', ``relate'') of existing scientific and technological knowledge and artifacts. Taken together, these patterns suggest a substantive shift in science and technology over time, with discovery and invention becoming less disruptive in nature, consistent with our results using the CD index.

The aggregate trends we document mask considerable heterogeneity in the disruptiveness of individual papers and patents and remarkable stability in the absolute number of highly disruptive works (see Methods, Extended Data Figure 5). Specifically, despite large increases in scientific productivity, the number of papers and patents with $CD_5$ values in the far right tail of the distribution remains nearly constant over time. This ``conservation'' of the absolute number of highly disruptive papers and patents holds despite considerable churn in the underlying fields responsible for producing those works (see Extended Data Figure 5, inset). These results suggest that the persistence of major breakthroughs---e.g., measurement of gravity waves, mRNA COVID-19 vaccines---is not inconsistent with slowing innovative activity. In short, declining aggregate disruptiveness does not preclude individually highly disruptive works.

What is driving the decline in disruptiveness? Earlier, we suggested our results are not consistent with explanations that link slowing innovative activity to diminishing ``low-hanging fruit.'' Extended Data Figure 6 further shows that the decline in disruptiveness is unlikely due to other field-specific factors by decomposing variation in $CD_5$ attributable to field, author, and year effects (see Methods).

Declining rates of disruptive activity are unlikely caused by the diminishing quality of science and technology \autocite{jaffe2011innovation,ioannidis2005most}. If they were, then the patterns seen in Figure 2 should be less visible in high quality work. However, when we restrict our sample to articles published in premier publication venues like \textit{Nature}, \textit{PNAS}, and \textit{Science} or to Nobel-winning discoveries (Extended Data Figure 7)\autocite{li2019dataset}, the downward trend persists. 

Furthermore, the trend is not driven by characteristics of the WoS and UPSTO data or our particular derivation of the $CD_5$ index; we observe similar declines in disruptiveness when we compute the $CD_5$ index on papers in JSTOR, the American Physical Society corpus, Microsoft Academic Graph, and PubMed (see Methods), the results of which are shown in Extended Data Figure 8. We further show that the decline is not an artifact of the CD index by reporting similar patterns using alternative derivations\autocite{bornmann2020disruption,leydesdorff2021proposal} (see Methods, Extended Data Figure 9).

Declines in disruptiveness are also not attributable to changing publication, citation, or authorship practices (see Methods). First, using approaches from the bibliometrics literature \autocite{bornmann2015methods, waltman2019field, waltman2016review, bornmann2020can, petersen2019methods}, we computed several normalized versions of the $CD_5$ index that adjusted for the increasing tendency for papers and patents to cite prior work\autocite{bornmann2015growth,bornmann2021growth}. Results using these alternative indicators (see Extended Data Figure 10) were similar to those we reported in Figure 2. Second, using regression, we estimated models of the $CD_5$ index as a function of indicator variables for each paper or patent's publication year, along with specific controls for field × year level—number of new papers/patents, mean number of papers/patents cited, mean number of authors/inventors per paper—and paper/patent-level—number of papers/patents cited—factors. Predictions from these models indicated a decline in disruptive papers and patents (see Extended Data Figure 11) that was consistent with our main results. Finally, using Monte Carlo simulations, we randomly rewired the observed citation networks while preserving key characteristics of scientists' and inventors' citation behavior, including the number of citations made and received by individual papers and patents and the age gap between citing and cited works. We find that observed $CD_5$ values are lower than those from the simulated networks (Extended Data Figure 12), and the gap is widening: over time, papers and patents are increasingly less disruptive than would be expected by chance. Taken together, these additional analyses indicate that the decline in $CD_5$ is unlikely to be driven by changing publication, citation, or authorship practices.

We also considered how declining disruptiveness relates to the growth of knowledge (Extended Data Figure 13). On the one hand, scientists and inventors face an increasing knowledge burden, which may inhibit discoveries and inventions that disrupt the status quo. On the other hand, as previously noted, philosophers of science suggest that existing knowledge fosters discovery and invention \autocite{koyre1952unpublished,weitzman1998recombinant,acemoglu2016innovation}. Using regression models, we evaluated the relationship between the stock of papers and patents (a proxy for knowledge) within fields and their $CD_5$ (Supplementary Table 1). Interestingly, we find a positive effect of the growth of knowledge on disruptiveness for papers, consistent with prior work \autocite{chu2021}; however, we find a negative effect for patents.

Given these conflicting results, we considered the possibility that the availability of knowledge may differ from its use. In particular, the growth in publishing and patenting may lead scientists and inventors to focus on narrower slices of prior work \autocite{jones2009burden,jones2011age}, thereby limiting the ``effective'' stock of knowledge. Using three proxies, we document a decline in the use of prior knowledge among scientists and inventors (Figure 4). First, we see a decline in the diversity of work cited (Figure 4A and D), indicating that contemporary science and technology are engaging with narrower slices of existing knowledge. Moreover, this decline in diversity is accompanied by an increase in the share of citations to the 1\% most highly cited papers and patents (Figures 4A1 and 4D1), which are also decreasing in semantic diversity (Figures 4A2 and 4D2). Over time, scientists and inventors are increasingly citing the same prior work, and that prior work is becoming more topically similar. Second, we see an increase in self-citation (Figure 4B and E), a common proxy for the continuation of one's pre-existing research stream \autocite{bonzi1991motivations,fowler2007does,king2017men}, which is consistent with scientists and inventors relying more on highly familiar knowledge. Third, the mean age of work cited, a common measure for the use of dated knowledge \autocite{mukherjee2017nearly,merton1961singletons,wang2013quantifying}, is increasing (Figure 4C and F) suggesting that scientists and inventors may be struggling to keep up with the pace of knowledge expansion and instead relying on older, familiar work. All three indicators point to a consistent story: a narrower scope of existing knowledge is informing contemporary discovery and invention. 

Results from a subsequent series of regression models suggest that use of less diverse work, more of one's own work, and older work are all negatively associated with disruption (see Methods, Extended Data Table 2), a pattern that holds even after accounting for the average age and number of prior works produced by team members. When the range of work used by scientists and inventors narrows, disruptive activity declines.

In summary, we report a dramatic decline in disruptive science and technology over time. Our analyses show that this trend is unlikely driven by changes in citation practices or the quality of published work. Rather, the decline represents a substantive shift in science and technology, one that reinforces concerns about slowing innovative activity. We further show that the documented trend is attributable at least in part to scientists' and inventors' reliance on a narrower set of existing knowledge. While philosophers of science may be correct that the growth of knowledge is an endogenous process---wherein accumulated understanding promotes future discovery and invention—for that process to play out, scientists and inventors need to engage with what is already known, a requirement that appears more difficult with time. 

As we noted at the beginning of this article, the past century witnessed unprecedented progress in science and technology. But what about the next 100 years? Our results offer some reasons for optimism. Notwithstanding aggregate declines, the absolute number of highly disruptive papers and patents produced is remarkably stable. Thus, science and technology do not necessarily seem to have reached the end of the ``endless frontier;'' rather, the decline may be driven instead by changing practices of scientific and technological production. The dramatic growth in the number of papers and patents produced annually, coupled with declining use of prior knowledge, suggest, for instance, that scientists and inventors may be allocating too much effort to writing and not enough to reading (and thinking). If so, the trend of declining disruptive science and technology may be reversible, likely with deeper understanding of the problem and rethinking of strategies for organizing the production of science and technology in the future.

\pagebreak

\newgeometry{top=0.5in, bottom=1in, left=0.5in, right=0.5in}

\begin{landscape}
\begin{figure}[H]
\includegraphics[width=\textwidth]{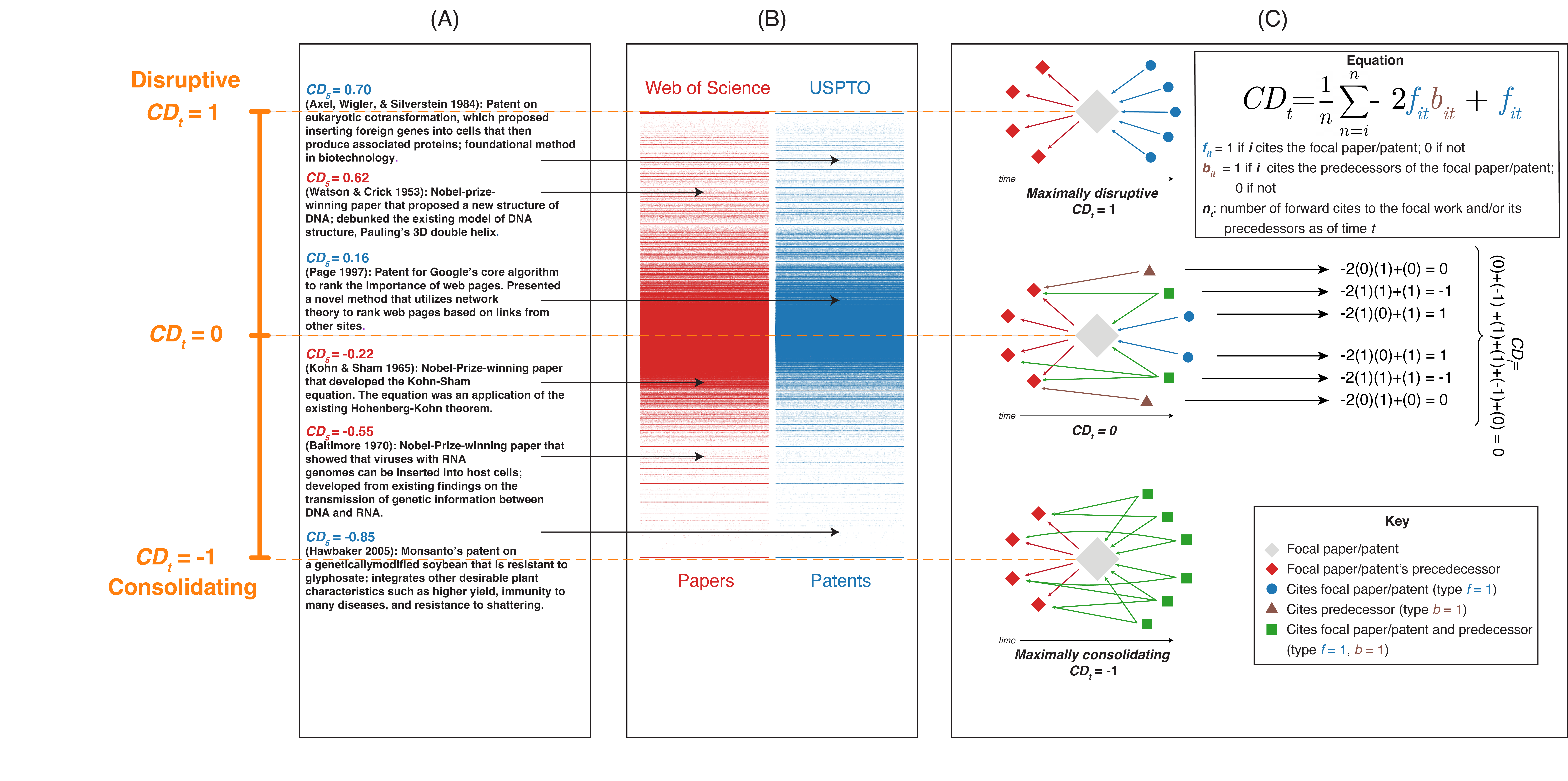}
\caption{\textbf{Overview of the measurement approach.} The figure below presents a schematic visualization of the CD index. \textbf{A} shows the $CD_5$ measure of three Nobel-prize-winning papers and three notable patents in our sample. \textbf{B} shows the distribution of the $CD_5$ measure for 24,962,339 papers form WoS between 1945-2010 and 3,434,055 patents from USPTO between 1976-2010, where a single dot represents a paper or patent. The vertical (up/down) dimension of each ``strip'' corresponds to values of the CD index (with axis values shown in orange on the left). The horizontal (left/right) dimension on each strip is included to minimize overlapping of the points. Darker areas on each strip plot indicate denser regions of the distribution (i.e., more common empirically observed values of the $CD_5$ index). Additional details on the distribution of the CD index are given in Extended Data Figure 1. \textbf{C} shows three hypothetical citation networks where the CD index is at its the maximally disruptive value ($CD_t$ = 1), midpoint value ($CD_t$ = 0), and maximally consolidating value ($CD_t$ = -1). The panel also provides the equation of the CD index and an illustrative calculation.}
\label{figure:cdschematic}
\end{figure}
\end{landscape}

\pagebreak

\begin{figure}[H]
\includegraphics[width=\textwidth]{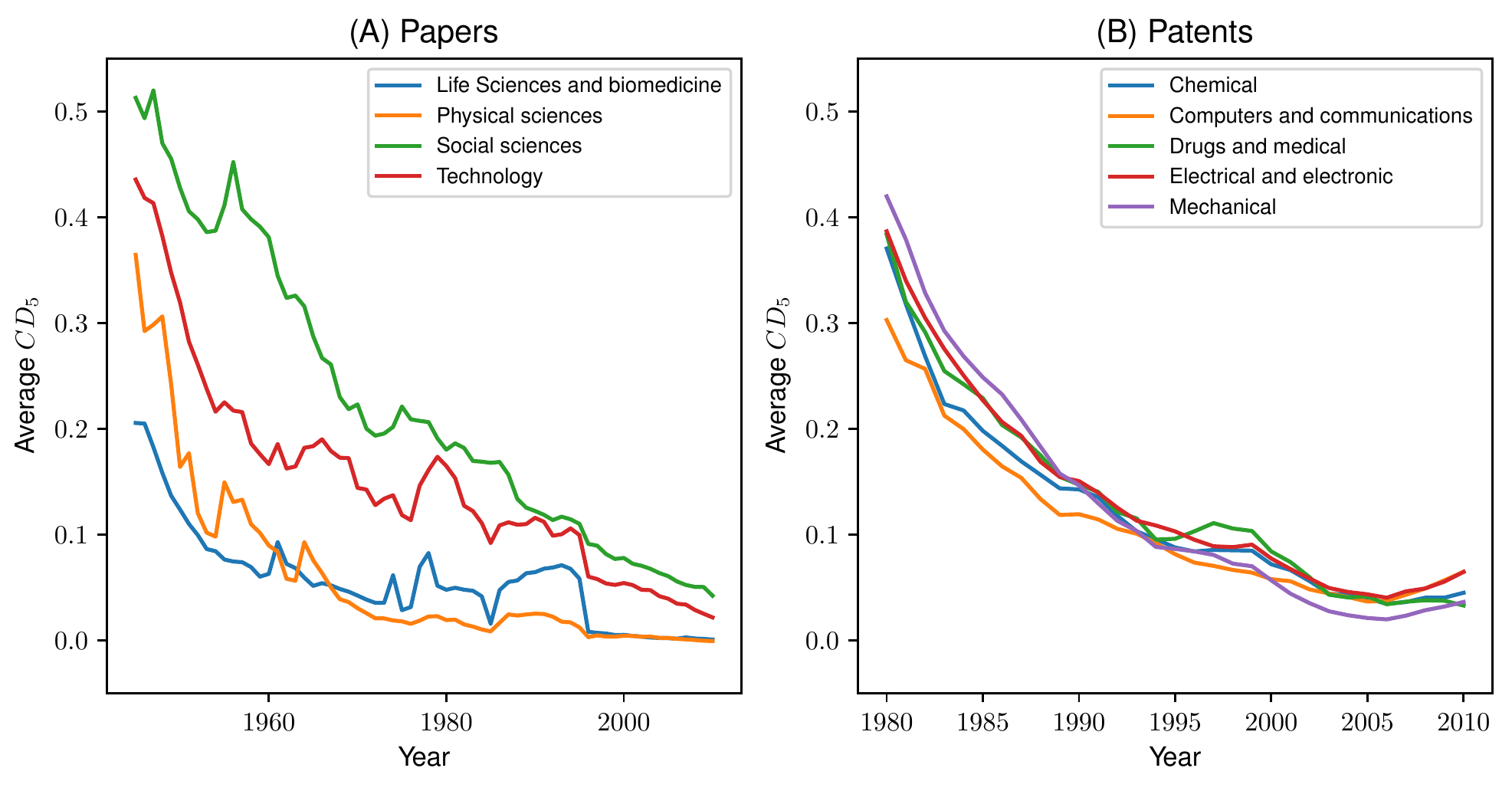}
\caption{\textbf{Decline of disruptive science and technology.} This figure shows the decline in $CD_5$ over time, separately for WoS papers (\textbf{A}) and USPTO patents (\textbf{B}). For papers, lines correspond to WoS research areas; from 1945-2010 the magnitude of decline ranges from 91.9\% (Social Science) to 100\% (Physical Science). For patents, lines correspond to NBER technology categories; from 1980-2010 the magnitude of decline ranges from 93.5\% (Computers and Communications) to 96.4\% (Drugs and Medical). As we elaborate in the Methods, this pattern of decline is robust to adjustment for confounding from changes in publication, citation, and authorship practices over time.  }
\label{figure:CDOvertime1}
\end{figure}

\pagebreak

\begin{landscape}
\begin{figure}[H]
\includegraphics[width=0.9\textwidth]{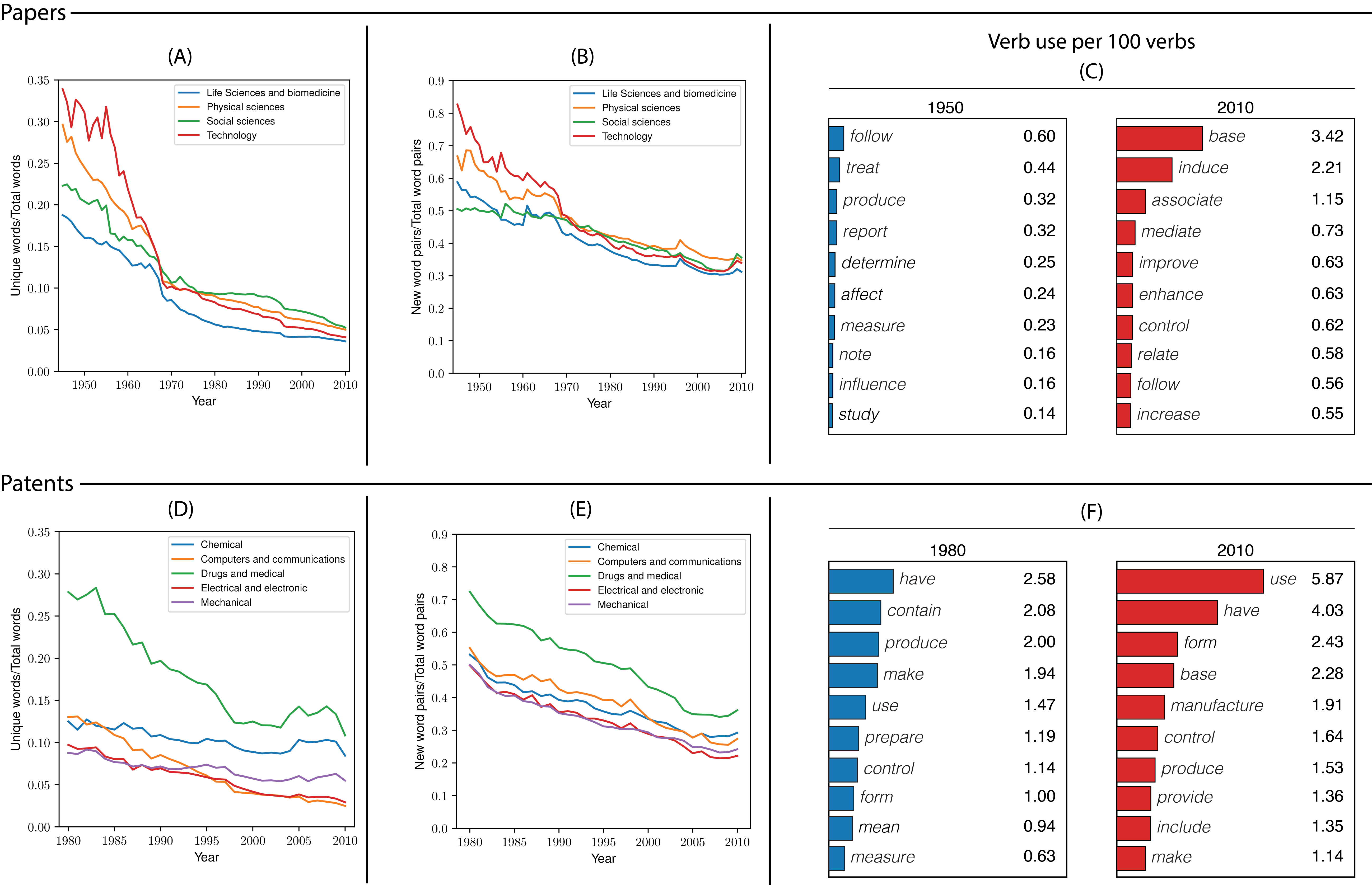}
\caption{\textbf{The decline of disruptive science and technology is visible in the changing language of papers and patents.} \textbf{A} and \textbf{D} show a decline in the diversity of language used in science and technology based on the unique/total words of WoS paper titles from 1945-2010 (\textbf{A}) and in USPTO patent titles from 1980-2010 (\textbf{D}). \textbf{B} and \textbf{E} show a decline in the novelty of language used in science and technology based on the number of new word pairs/total word pairs introduced each year in WoS paper titles from 1945-2010 (\textbf{B}) and in USPTO patent titles from 1980-2010 (\textbf{E}). For papers in both \textbf{A} and \textbf{B}, lines correspond to WoS research areas. For patents in both \textbf{D} and \textbf{E}, lines correspond to NBER technology categories. \textbf{C} and \textbf{F} show the frequency of the most commonly used verbs in paper titles for the first (blue) and last (red) decades of the observation period in paper (\textbf{C}) and patent titles (\textbf{F}).}
\label{figure:LexicalDiversityTitle}
\end{figure}
\end{landscape}

\pagebreak

\begin{landscape}
\begin{figure}[H]
\includegraphics[width=0.9\textwidth]{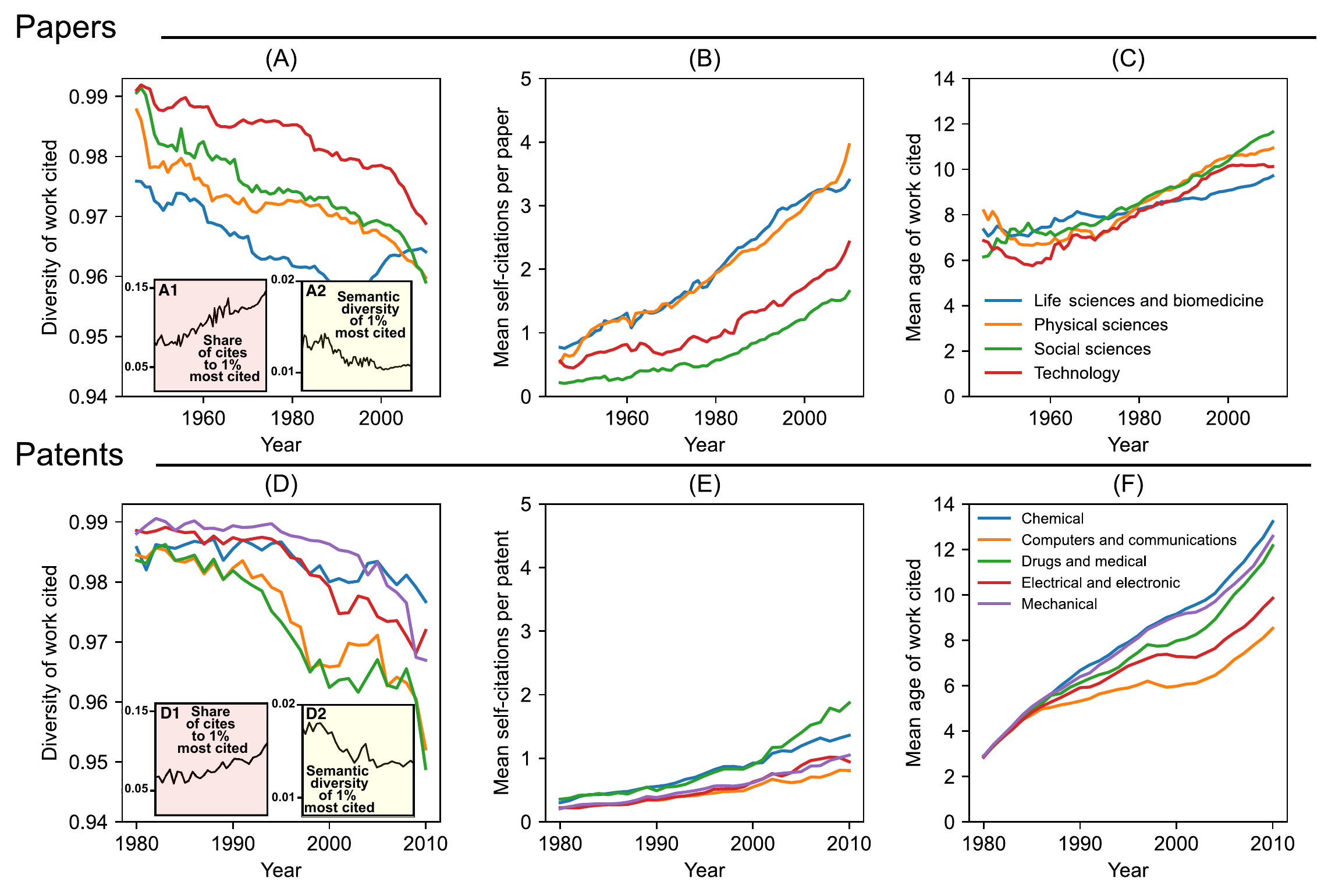}
\caption{\textbf{Papers and patents are using narrower portions of existing knowledge.} This figure shows changes in the level of diversity of existing scientific and technological knowledge use in papers (\textbf{A}, \textbf{B}, and \textbf{C}) and patents (\textbf{D}, \textbf{E}, and \textbf{F}) based on following measures: diversity of work cited (\textbf{A} and \textbf{D}), mean number of self-citations (\textbf{B} and \textbf{E}), and mean age of cited work (\textbf{C} and \textbf{F}). The inset plots of \textbf{A} and \textbf{D} show changes in the share of citations to the top 1\% most highly cited papers (\textbf{A1} and \textbf{D1}) and in the semantic diversity of the top 1\% most cited over time (\textbf{A2} and \textbf{D2}). Values of both measures are computed within field and year, and are subsequently averaged across fields for plotting. Semantic diversity is based on paper and patent titles; values correspond to the ratio of the standard deviation to the mean pairwise cosine similarity (i.e., the coefficient of variation) among the titles of the 1\% most cited papers and patents by field and year. To enable semantic comparisons, titles were vectorized using pretrained word embeddings. For papers, lines are shown for each WoS research area; for patents, lines are shown for each NBER technology category. In subsequent regression analyses using these measures, we find that using less diverse work, more of one's own work, and older work is associated with less disruptive papers and patents (see Methods, Extended Data Table 2).}
\label{figure:utilization}
\end{figure}
\end{landscape}

\restoregeometry

\pagebreak
\printbibliography

\pagebreak

\section*{Methods}
\begin{refsection}

Here, we provide details of the WoS and the USPTO datasets, the two main corpora analyzed in this paper. Subsequently, we provide additional information about the specific analyses conducted. 

\subparagraph{WoS dataset} We limit our focus to research papers published between 1945 and 2010. Although the WoS data begins in the year 1900, the scale and social organization of science shifted dramatically in the post-war era, thereby making comparisons with the present difficult and potentially misleading \autocite{guimera2005team,jones2008multi,wuchty2007increasing}. We end our analyses of papers in 2010 because some of our measures require several subsequent years of data following paper publication. The WoS data archives 64 million documents published in 28,968 journals between 1900 to 2017 and 224 million citations among them. In addition, the WoS data includes titles and the full text of abstracts for 28 million records published between 1913 to 2017.

\subparagraph{USPTO dataset} We limit our focus to patents granted from 1976, which is the earliest year for which machine-readable records are available in the USPTO data. As we did with papers, we end our analyses in 2010 because some measures require data from subsequent years for calculation. The USPTO data is the most exhaustive source of historical data on inventions, with information on 7 million patents granted between 1976 and 2017 and their corresponding 9 million citations. The USPTO data includes the titles and abstracts for 6 million patents granted between 1976 and 2017.

\subparagraph{Analysis of highly disruptive papers and patents} Observations (and claims) of slowing progress in science and technology are increasingly common, supported not just by the evidence we report, but also by prior research from diverse methodological and disciplinary perspectives\autocite{jones2009burden,gordon2016rise,chu2021,pammolli2011productivity,bloom2020ideas,packalen2020nih, horgan2015end,collison2018science,jaffe2011innovation}. Yet as noted in the main text, there is a tension between observations of slowing progress from aggregate data on the one hand, and continuing reports of seemingly major breakthroughs in many fields of science and technology---spanning everything from the measurement of gravity waves to the sequencing of the human genome---on the other. In an effort to reconcile this tension, we considered the possibility that while overall, discovery and invention may be less disruptive over time, the high-level view taken in prior work may mask considerable heterogeneity. Put differently, aggregate evidence of slowing progress does not preclude the possibility that some (smaller) subset of discoveries and inventions are highly disruptive. 

To evaluate this possibility, we plot the number of disruptive papers (Extended Data Figure 5A) and patents (Extended Data Figure 5B) over time, where disruptive papers and patents are defined as those with $CD_5$ values $>$ 0. Within each panel, we plot four lines, corresponding to four evenly spaced intervals—(0,0, 0.25], (0.25, 0.5], (0.5, 0.75], (0.75, 1.00]—over the positive values of the $CD_5$ index. The first two intervals therefore correspond to papers and patents that are relatively weakly disruptive, while the latter two correspond to those that are more strongly so (e.g., where we may expect to see major breakthroughs like some of those mentioned above). Despite major increases in the numbers of papers and patents published each year, we see little change in the number of highly disruptive papers and patents, as evidenced by the relatively flat red, green, and orange lines. Remarkably, this ``conservation'' of disruptive work holds even despite fluctuations over time in the composition of the scientific and technological fields responsible for producing the most disruptive work (see Extended Data Figure 5, inset plots). Overall, these results help to account for simultaneous observations of both major breakthroughs in many fields of science and technology and aggregate evidence of slowing progress.  

\subparagraph{Analysis of the relative contribution of field, year, and author/inventor effects} Our results show a steady decline in the disruptiveness of science and technology over time. Moreover, the patterns we observe are generally similar across broad fields of study, which suggests that the factors driving the decline are not unique to specific domains of science and technology. The decline could be driven by other factors, such as the conditions of science and technology at a point in time or the particular individuals that produce science and technology. For example, exogenous factors like economic conditions may encourage research or invention practices that are less disruptive. Similarly, scientists and inventors of different generations may have different approaches, which may result in greater or lesser tendencies for producing disruptive work. We therefore sought to understand the relative contribution of field, year, and author (or inventor) factors to the decline of disruptive in science and technology. 

To do so, we decomposed the relative contribution of field, year, and author fixed effects to the predictive power of regression models of the CD index. The unit of observation in these regressions is the author (or inventor) $\times$ year. We enter field fixed effects using granular subfield indicators (i.e., 150 WoS subjects for papers, 138 NBER subcategory for patents). For simplicity, we did not include additional covariates beyond the fixed effects in our models. Field fixed effects capture all field-specific factors that do not vary by author or year (e.g., the basic subject matter); year fixed effects capture all year-specific factors that do not vary by field or author (e.g., the state of communication technology); author (or inventor) fixed effects capture all author-specific factors that do not vary by field or year (e.g., the year of PhD awarding). After specifying our model, we determine the relative contribution of field, year, and author fixed effects to the overall model adjusted $R^2$ using Shapley-Owen decomposition. Specifically, given our $n=3$ groups of fixed effects (field, year, and author) we evaluate the relative contribution of each set of fixed effects by estimating the adjusted $R^2$ separately for the $2^n$ models using subsets of the predictors. The relative contribution of each set of fixed effects is then computed using the Shapley value from game theory \autocite{gromping2007estimators}.

Results of this analysis are shown in Extended Data Figure 6, for both papers (top bar) and patents (bottom bar). Total bar size corresponds to the value of the adjusted $R^2$ for the fully specified model (i.e., with all three groups of fixed effects). Consistent with our observations from plots of the CD index over time, we observe that for both papers and patents, field specific factors make the lowest relative contribution to the adjusted $R^2$ (0.02 and 0.01 for papers and patents, respectively). Author fixed effects, by contrast, appear to contribute much more to the predictive power of the model, for both papers (0.20) and patents (0.17). Researchers and inventors who enter the field in more recent years may face a higher burden of knowledge and thus resort to building on narrower slices of existing work (e.g., due to more specialized doctoral training), which would generally lead to less disruptive science and technology being produced in later years, consistent with our findings. The pattern is more complex for year fixed effects; while year-specific factors that do not vary by field or author hold more explanatory power than field for both papers (0.01) and patents (0.16), they appear to be substantially more important for the latter than the former. Taken together, these findings suggest that relatively stable factors that vary across individual scientists and inventors may be particularly important for understanding changes in disruptiveness over time. The results also confirm that domain-specific factors across fields of science and technology play a very small role in explaining the decline in disruptiveness of papers and patents.

\subparagraph{Alternative samples} We also considered whether the patterns we document may be artifacts of our choice of data sources. While we observe consistent trends in both the WoS and USPTO data, and both databases are widely used by the Science of Science community, our results may conceivably driven by factors like changes in coverage (e.g., journals added or excluded from WoS over time) or even data errors rather than fundamental changes in science and technology. To evaluate this possibility, we therefore calculated the $CD_5$ index for papers in four additional databases---JSTOR, the American Physical Society corpus, Microsoft Academic Graph, and PubMed. We included all records from 1930-2010 from PubMed (16,774,282 papers), JSTOR (1,703,353 papers), and American Physical Society (478,373 papers).\footnote{The JSTOR data were obtained via a special request from ITHAKA, the data maintainer (\url{http://www.ithaka.org}), as were the American Physical Society data (see \url{https://journals.aps.org/datasets}). We downloaded the Microsoft Academic Graph data from CADRE at Indiana University (\url{https://cadre.iu.edu/}). The PubMed data were downloaded from the National Library of Medicine FTP server (\url{ftp://ftp.ncbi.nlm.nih.gov/pubmed/baseline}).} Due to the exceptionally large scale of Microsoft Academic Graph and the associated computational burden, we randomly extracted 1 million papers. As shown in Extended Data Figure 8, the downward trend in disruptiveness is evident across all samples.

\subparagraph{Alternative bibliometric measures} 
Several recent papers have introduced alternative specifications of CD index \autocite{funk2017dynamic}. We evaluated whether the declines in disruptiveness we observe are corroborated using two alternative variations. One criticism of the CD index has been that the number of papers that cite only the focal paper's references dominates the measure\autocite{bornmann2020disruption}. Bornmann et al.\autocite{bornmann2020disruption} proposes $DI_{l}^{no\,k}$ as the key variant that is less susceptible this issue. Another potential weakness of the CD index is that it could be very sensitive to small changes in the forward citation patterns of papers that make no backward citations\autocite{leydesdorff2021proposal}. Leydesdorff et al.\autocite{leydesdorff2021proposal} suggests $DI^*$ as an alternate indicator of disruption that addresses this issue. Therefore, we calculated $DI_{l}^{no\,k}$ where $l = 1$\autocite{bornmann2020disruption} and $DI^*$\autocite{leydesdorff2021proposal} for 100,000 randomly drawn papers and patents each from our analytic sample. Results are presented in Extended Data Figure 9A (papers) and 9B (patents). The blue lines indicate disruption based on Bornmann et al.\autocite{bornmann2020disruption} and the orange lines indicate disruption based on Leydesdorff et al.\autocite{leydesdorff2021proposal}. Across science and technology, the two alternative measures both show declines in disruption over time, similar to the patterns observed with the CD index. Taken together, these results suggest that the declines in disruption we document are not an artifact of our particular operationalization.

\subparagraph{Robustness to changes in publication, citation, and authorship practices} We also considered whether our results may be attributable to changes in publication, citation, or authorship practices, rather than by substantive shifts in discovery and invention. Perhaps most critically, as noted in the main text, there has been a dramatic expansion in publishing and patenting over the period of our study. This expansion has naturally increased the amount of prior work that is relevant to current science and technology and therefore at risk of being cited, a pattern reflected in the dramatic increase in the average number of citations made by papers and patents (i.e., papers and patents are citing more prior work than in previous eras)\autocite{bornmann2015growth,bornmann2021growth}. Recall that the CD index quantifies the degree to which future work cites a focal work together with its predecessors (i.e., the references in the bibliography of the focal work). Greater citation of a focal work independently of its predecessors is taken to be evidence of a social process of disruption. As papers and patents cite more prior work, however, the probability of a focal work being cited independently of its predecessors may decline mechanically; the more citations a focal work makes, the more likely future work is to cite it together with one of its predecessors, even by chance. Consequently, increases in the number of papers and patents available for citing and in the average number of citations made by scientists and inventors may contribute to the declining values of the CD index. In short, given the dramatic changes in science and technology over our long study window, the CD index of papers and patents published in earlier periods may not be directly comparable to those of more recent vintage, which could in turn render our conclusions about the decline in disruptive science and technology suspect. We addressed these concerns using three distinctive but complementary approaches---normalization, regression adjustment, and simulation.

\emph{Verification using normalization.} First, following common practice in bibliometric research\autocite{bornmann2015methods, waltman2019field, waltman2016review, bornmann2020can, petersen2019methods}, we developed two normalized versions of the CD index, with the goal of facilitating comparisons across time. Among the various components of the CD index, we focused our attention on the count of papers or patents that only cite the focal work's references (``$N_k$''), as this term would seem most likely to scale with the increases in publishing and patenting and in the average number of citations made by papers and patents to prior work \autocite{bornmann2020disruption}. Larger values of $N_k$ lead to smaller values of the CD index. Consequently, dramatic increases in $N_k$ over time, particularly relative to other components of the measure, may lead to a downward bias, thereby inhibiting our ability to accurately compare disruptive science and technology in later years with earlier periods.

Our two normalized versions of the CD index aim to address this potential bias by attenuating the effect of increases in $N_k$. In the first version, which we call ``Paper normalized,'' we subtract from $N_k$ the number of citations made by the focal paper or patent to prior work (``$N_b$''). The intuition behind this adjustment is that when a focal paper or patent cites more prior work, $N_k$ is likely to be larger because there are more opportunities for future work to cite the focal paper or patent's predecessors. This increase in $N_k$ would result in lower values of the CD index, though not necessarily as a result of the focal paper or patent being less disruptive. In the second version, which we call ``Field $\times$ year normalized,'' we subtract $N_k$ by the average number of backward citations made by papers or patents in the focal paper or patent's WoS research area or NBER technology category, respectively, during its year of publication (we label this quantity ``$N_{b}^{mean}$''). The intuition behind this adjustment is that in fields and time periods in which there is a greater tendency for scientists and inventors to cite prior work, $N_k$ is also likely to be larger, thereby leading to lower values of the CD index, though again not necessarily as a result of the focal paper or patent being less disruptive. In cases where either $N_b$ or $N_{b}^{mean}$ exceed the value of $N_k$, we set $N_k$ to 0 (i.e., $N_k$ is never negative in the normalized measures). Both adaptations of the CD index are inspired by established approaches in the scientometrics literature, and may be understood as a form of ``citing side normalization'' (i.e., normalization by correcting for the effect of differences in lengths of references lists \autocite{waltman2016review}).

In Extended Data Figure 10, we plot the average values of both normalized versions of the CD index over time, separately for papers (Panel A) and Patents (Panel B). Consistent with our findings reported in the main text, we continue to observe a decline in the CD index over time, suggesting that the patterns we observe in disruptive science and technology are unlikely driven changes in citation practices.


\emph{Verification using regression adjustment.} Second, we adjusted for potential confounding using a regression-based approach. This approach complements the bibliometric normalizations just described by allowing us to account for a broader array of changes in publication, citation, and authorship in general (the latter of which is not directly accounted for in either the normalization approach or the simulation approach described next), and increases the amount of prior work that is relevant to current science and technology in particular. In Supplementary Table 2, we report the results of regression models predicting the $CD_5$ index of papers (Models 1-4) and patents (Models 5-8), with indicator variables included for each year of our study window (the reference categories are 1945 and 1980 for papers and patents, respectively). Models 1 and 4 are the baseline models, and include no other adjustments beyond the year indicators. In Models 2 and 5, we add subfield fixed effects (Web of Science subject areas for papers, NBER technology subcategories for patents). Finally, in Models 3-4 and 7-8, we add control variables for several field $\times$ year level---\emph{Number of new papers/patents}, \emph{Mean number of papers/patents cited}, \emph{Mean number of authors/inventors per paper}---and paper/patent-level---\emph{Number of papers/patents cited}---characteristics, thereby enabling more robust comparisons in patterns of disruptive science and technology over the long time period spanned by our study. We find that the inclusion of these controls improves model fit, as indicated  by statistically significant Wald tests presented below the relevant models.

Across all 8 models shown in Supplementary Table 2, we find that the coefficients on the year indicators are statistically significant and negative, and growing in magnitude over time, which is consistent with the patterns we reported based on unadjusted values on the $CD_5$ index in the main text (i.e., Extended Data Figure 11). In Extended Data Figure 11, we visualize the results of our regression-based approach by plotting the predicted values of the $CD_5$ index separately for each of the year indicators included in Models 4 (papers) and 8 (patents). To enable comparisons with the raw values of the $CD_5$ index shown in the main text, we present the separate predictions made for each year as a line graph. As shown in the figure, we continue to observe declining values of the CD index across papers and patents, even when accounting for changes in publication, citation, and authorship practices.

\emph{Verification using simulation.} Third, following related work in the Science of Science\autocite{uzzi2013atypical, mukherjee2016new, christianson2020architecture, newman2001structure}, we considered whether our results may be an artifact of changing patterns in publishing and citation practices by using a simulation approach. In essence, the CD index measures disruption by characterizing the network of citations around a focal paper or patent. However, many complex networks, even those resulting from random processes, exhibit structures that yield nontrivial values on common network measures (e.g., clustering)\autocite{uzzi2005collaboration, newman2001scientific, funk2014making}. During the period spanned by our study, the citation networks of science and technology experienced significant change, with dramatic increases in both the numbers of nodes (i.e., papers or patents) and edges (i.e., citations). Thus, rather than reflecting a meaningful social process, the observed declines in disruption may result from these structural changes in the underlying citation networks.    

To evaluate this possibility, we followed standard techniques from network science\autocite{barabasi2016network, uzzi2005collaboration} and conducted an analysis in which we recomputed the $CD_5$ index on randomly rewired citation networks. If the patterns we observe in the $CD_5$ index are the result of structural changes in the citation networks of science and technology (e.g., growth in the number of nodes or edges) rather than a meaningful social process, then these patterns should also be visible in comparable random networks that experience similar structural changes. Therefore, finding that the patterns we see in the $CD_5$ index differ for the observed and random citation networks would serve as evidence that the decline in disruption is not an artifact of the data. 

We began by creating copies of the underlying citation network on which the values of the $CD_5$ index used in all analyses reported in the main text were based, separately for papers and patents. For each citation network (one for papers, one for patents), we then rewired citations using a degree-preserving randomization algorithm. In each iteration of the algorithm, two edges (e.g., A-B, C-D) are selected from the underlying citation network, after which the algorithm attempts to swap the two endpoints of the edges (e.g., A-B becomes A-D, and C-D becomes C-B). If the degree centrality of A, B, C, and D remains the same after the swap, the swap is retained; otherwise, the algorithm discards the swap and moves on to the next iteration. When evaluating degree centrality, we consider ``in-degree'' (i.e., citations from other papers/patents to the focal paper/patent) and ``out-degree'' (i.e., citations from the focal paper/patent to other papers/patents) separately. Furthermore, we also required that the age distribution of citing and cited papers/patents was identical in the original and rewired networks. Specifically, swaps were only retained when the publication year of the original and candidate citations were the same. In light of these design choices, our rewiring algorithm should be seen as fairly conservative, as it preserves substantial structure from the original network. There is no scholarly consensus on the number of swaps necessary to ensure the original and rewired networks are sufficiently different from one another; the rule we adopt here is $100\times m$, where $m$ is the number of edges in the network being rewired. 

Following previous work\autocite{uzzi2013atypical}, we created 10 rewired copies of the observed citation networks for both papers and patents. After creating these rewired citation networks, we then recomputed the $CD_5$ index. Due to the large scale of the WoS data, we base our analyses on a 10\% random subsample of papers; the $CD_5$ index was computed on the rewired network for all patents. For each paper and patent, we then compute a z-score that compares the observed value of the $CD_5$ index to those of the same paper or patent in the 10 rewired citation networks. Positive z-scores indicate that the observed $CD_5$ index value is greater (i.e., more disruptive) than would be expected by chance; negative z-scores indicate that the observed values are lesser (i.e., more consolidating). 

The results of these analyses are shown in Extended Data Figure 12, separately for papers (Extended Data Figure 12A) and patents (Extended Data Figure 12B). Lines correspond to the average z-score among papers or patents published in the focal year. The plots reveal a pattern of change in the $CD_5$ index over and beyond that ``baked in'' to the changing structure of the network. We find that on average, papers and patents tend to be less disruptive than would be expected by chance, and moreover, the gap between the observed $CD_5$ index values and those from the randomly rewired networks is increasing over time, which is consistent with our findings of a decline in disruptive science and technology. 

Taken together, the results of the foregoing analyses suggest that while there have been dramatic changes in science and technology over the course of our long study window, particularly with respect to publication, citation, and authorship practices, the decline in disruptive science and technology that we document using the CD index is unlikely an artifact of these changes, and instead represents a substantive shift in the nature of discovery and invention.

\subparagraph{Regression analysis of disruptiveness and the use of prior knowledge} We evaluate the relationship between disruptiveness and the use of prior knowledge using regression models, predicting $CD_5$ for individual papers and patents, based on three indicators of prior knowledge use---the diversity of work cited, mean number of self-citations, and mean age of work cited.\footnote{Our measure of the diversity of work cited is measured at the field $\times$ year level; all other variables included in the regressions are defined at the level of the paper or patent.} To account for potential confounding factors, our models included year and field fixed effects. Year fixed effects account for time variant factors that affect all observations (papers or patents) equally (e.g., global economic trends). Field fixed effects account for field-specific factors that do not change over time (e.g., some fields may intrinsically value disruptive work over consolidating ones). In contrast to our descriptive plots, for our regression models, we adjust for field effects using the more granular 150 WoS ``extended subjects''\footnote{E.g., ``Biochemistry \& Molecular Biology,'' ``Biophysics,'' ``Biotechnology \& Applied Microbiology,'' ``Cell Biology, Developmental Biology,'' ``Evolutionary Biology,'' and ``Microbiology'' are some of the extended subjects within the ``Life Sciences \& Biomedicine'' research area.}  and 38 NBER technology subcategories.\footnote{E.g., ``Agriculture, Food, Textile,'' ``Coating,'' ``Gas,'' ``Organic,'' and ``Resins'' are some of the subcategories within the ``Chemistry'' technology category.} 

In addition, we also include controls for the \emph{Mean age of team members}\footnote{We measure age as ``career age,'' defined as the difference between the publication year of the focal paper or patent and the first year in which each author or inventor published a paper or patent. For papers, we excluded a small number of cases (about 0.1\%) where the career age was unrealistically high (i.e., more than 80 years), and were likely to be the result of inaccuracies in the author-name disambiguation algorithm.} and the \emph{Mean number of prior works produced by team members}. While increases in rates of self citations may indicate that scientists and inventors are becoming more narrowly focused on their own work, these rates may also be driven in part by the amount of prior work available for self-citing. Similarly, while increases in the age of work cited in papers and patents may indicate that scientists and inventors are struggling to keep up, they may also be driven by the rapidly aging workforce in science and technology\autocite{blau2017us,cui2022aging}. For example, older scientists and inventors may be more familiar with or more attentive to older work, or may actively resist change\autocite{azoulay2019does}. These control variables help to account for these alternative explanations. 

Extended Data Table 1 shows summary statistics for variables used in the ordinary-least-squares regression models. The diversity of work cited is measured by normalized entropy, which ranges from 0 to 1. Greater values on this measure indicate a more uniform distribution of citations to a wider range of existing work; lower values indicate a more concentrated distribution of citations to a smaller range of existing work. The tables show that the normalized entropy in a given field and year has a nearly maximal average entropy of 0.98 for both science and technology. About 16\% of papers cited in a paper are by an author of the focal paper; the corresponding number for patents in about 7\%. Papers tend to rely on older work and work that varies more greatly in age (measured by standard deviation) than patents. Additionally, the average $CD_5$ of a paper is 0.04 while the average $CD_5$ of a patent is 0.12, meaning that the average paper tends to be less disruptive than the average patent.

We find that using more diverse work, less of one’s own work, and older work tends to be associated with the production of more disruptive science and technology, even after accounting for the average age and number of prior works produced by team members. These findings are based on our regression results, shown in Extended Data Table 2. Models 6 and 12 present the full regression models. The models indicate a consistent pattern for both science and technology, wherein  The coefficients for diversity of work cited are positive and significant for papers (0.159, $p < 0.01$) and patents (0.069, $p < 0.01$), indicating that in fields where there is more use of diverse work, there is greater disruption. Holding all other variables at their means, predicted $CD_5$ of papers and patents increase by 303.5\% and 1.3\%, respectively, when the diversity of work cited increases by one standard deviation. The coefficients of the ratio of self-citations to total work cited is negative and significant for papers (-0.011, $p < 0.01$) and patents (-0.060, $p < 0.01$), showing that when researchers or inventors rely more on their own work, discovery and invention tends to be less disruptive. Again holding all other variables at their means, the predicted $CD_5$ of papers and patents decrease by 622.9\% and 18.5\%, respectively, with a one standard deviation increase in the ratio. The coefficients of the interaction between mean age of work cited and dispersion in age of work cited is positive and significant for papers (0.000, $p < 0.01$) and patents (0.001, $p < 0.01$), suggesting that---holding the dispersion of the age of work cited constant---papers and patents that engage with older work are more likely to be disruptive. The predicted $CD_5$ of papers and patents increase by a striking 2,072.4\% and 58.4\%, respectively, when the mean age of work cited increases by one standard deviation (about 9 and 8 years for papers and patents, respectively), again holding all other variables at their means. In summary, the regression results suggest that changes in the use of prior knowledge may contribute to the production of less disruptive science and technology.

\pagebreak
\printbibliography[heading=subbibliography]
\end{refsection}

\setcounter{table}{0} 
\setcounter{figure}{0} 

\renewcommand{\figurename}{Extended Data Figure}
\renewcommand{\tablename}{Extended Data Table}

\pagebreak
\section*{Acknowledgments}
We thank the National Science Foundation for financial support of work related to this project (grants 1829168, 1932596, and 1829302). We would also like to thank Jonathan O. Allen, Thomas Gebhart, Daniel McFarland, Staša Milojević, Raviv Murciano-Goroff, and participants of the CADRE workshop at the Indiana University Network Science Institute for their feedback.

\section*{Author Contributions}
R.J.F. and E.L. collaboratively contributed to the conception and design of the study. R.J.F. and M.P. collaboratively contributed to the acquisition, analysis, and interpretation of the data. R.J.F. created software used in the study. R.J.F. E.L., and M.P. collaboratively drafted and revised the manuscript. 

\section*{Competing interests}
The authors declare no competing interests. 

\section*{Additional Information}

\noindent
\textbf{Supplementary Information} is available for this paper.

\noindent
\textbf{Correspondence and request for materials} should be addressed to R.J.F.

\pagebreak

\section*{Extended data}

\begin{figure}[H]
\includegraphics[width=\textwidth]{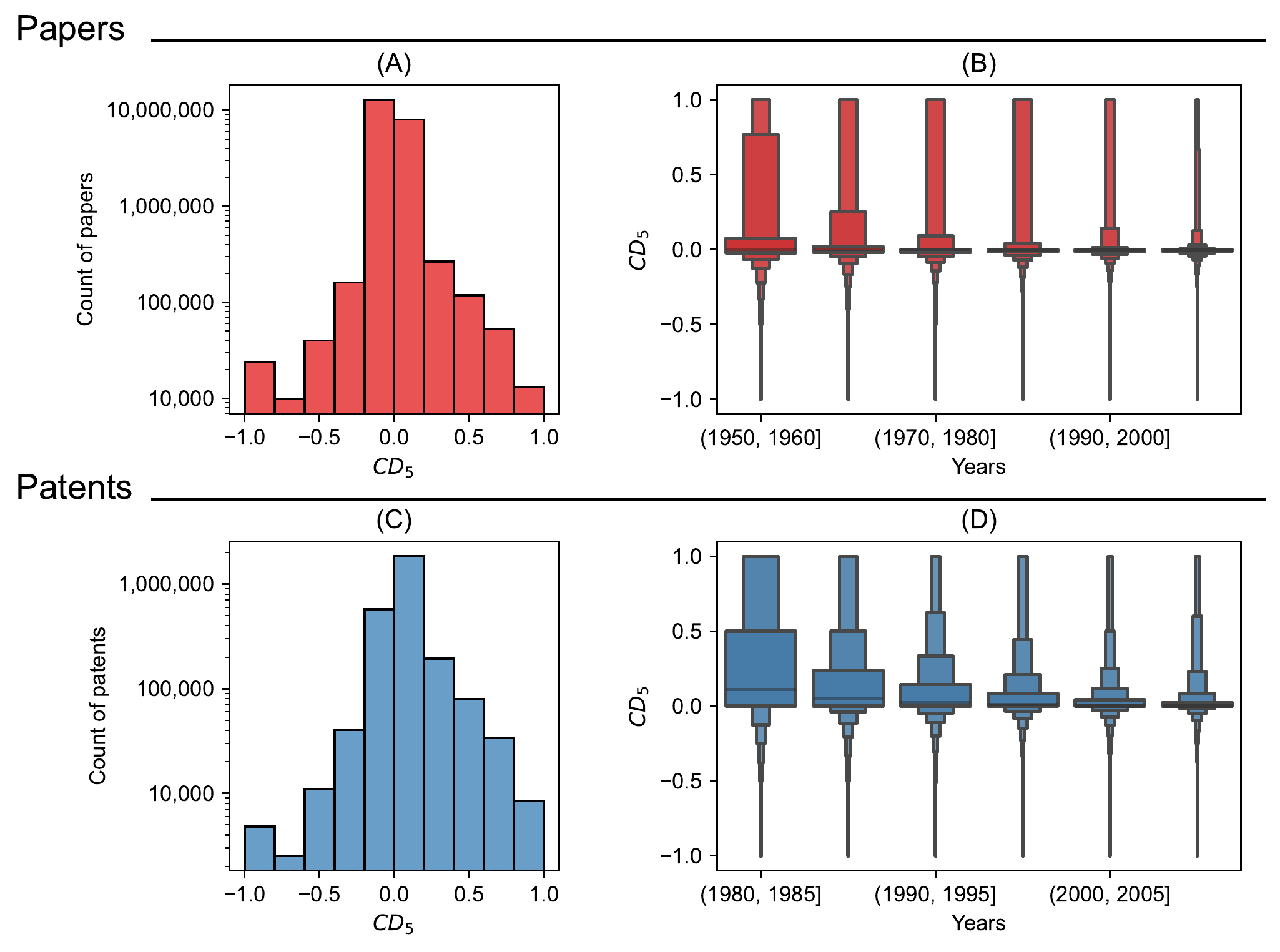}
\caption{\textbf{Distribution of $CD_5$.} This figure gives an overview of the distribution of $CD_5$ for papers and patents. Panels \textbf{A} and \textbf{C} show counts of papers and patents over discrete intervals of $CD_5$. Panels \textbf{B} and \textbf{D} show the distribution of $CD_5$ over time, within 10 (papers) and 5 (patents) year intervals, using letter-value plots\autocite{hofmann2017value}. These plots are similar to boxplots, but generally provide more reliable summaries for large datasets. They are drawn by identifying the median of the underlying distribution and then recursively drawing boxes outward from there in either direction that encompass half of the remaining data. }
\label{figure:CDDistribution}
\end{figure}
\pagebreak

\begin{figure}[H]
\includegraphics[width=\textwidth]{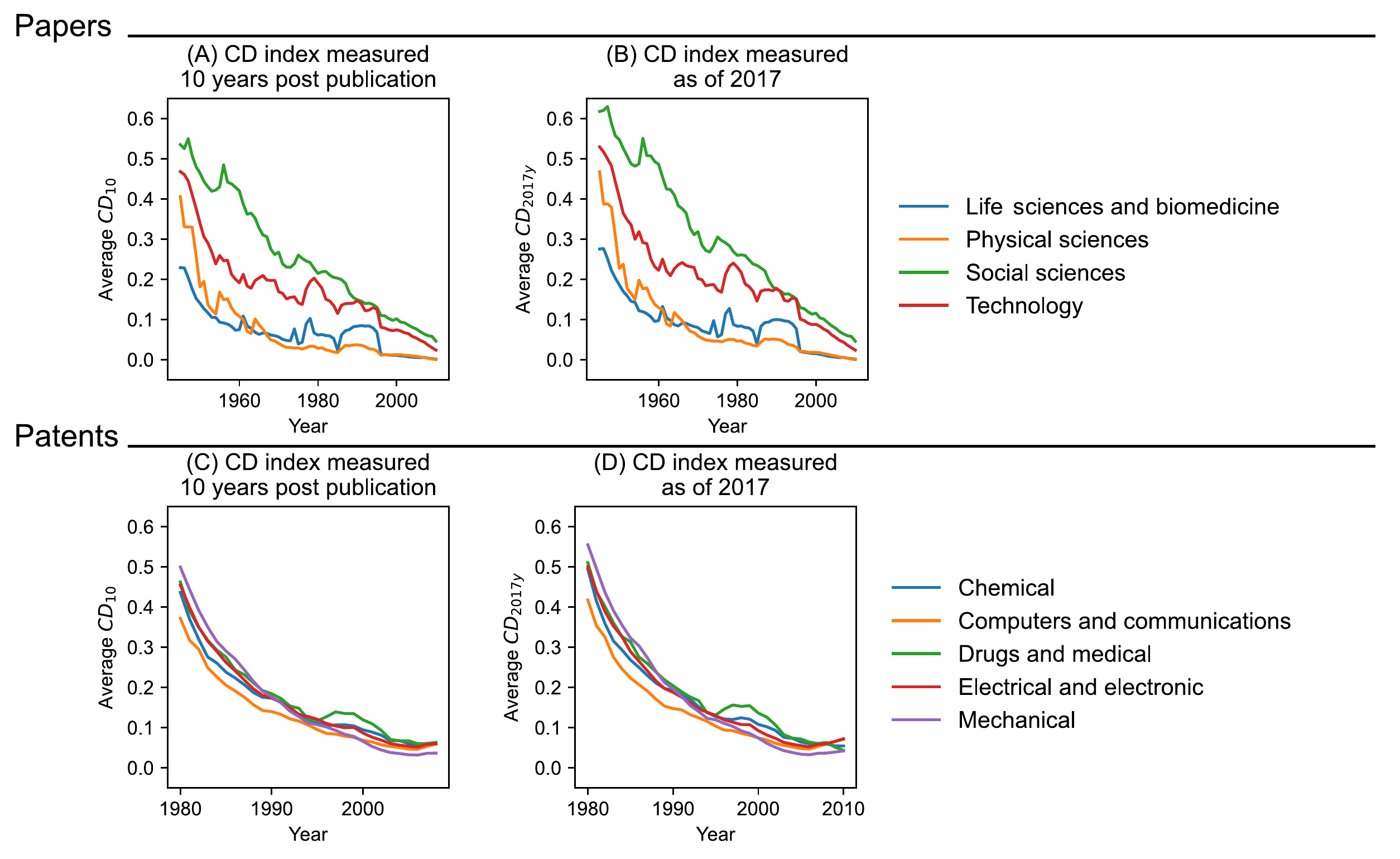}
\caption{\textbf{CD index measured using alternative forward citation windows.} This figure evaluates the sensitivity of our results to the use of different forward citation windows when computing the CD index. In the main text, the index is computed based on citations made to papers and patents and their backward references as of 5 years after the year of publication. \textbf{A} and \textbf{C} plot the CD index using a longer, 10 year forward window, for papers and patents, respectively. \textbf{B} and \textbf{D} plot the CD index using all forward citations made to sample papers and patents as of the year 2017. Overall, the results mirror those reported in the main text, although the decline is somewhat steeper using longer forward citation windows, suggesting our primary results may represent a more conservative estimate.}
\label{figure:CDAlternativeWindows}
\end{figure}
\pagebreak

\begin{figure}[H]
\includegraphics[width=\textwidth]{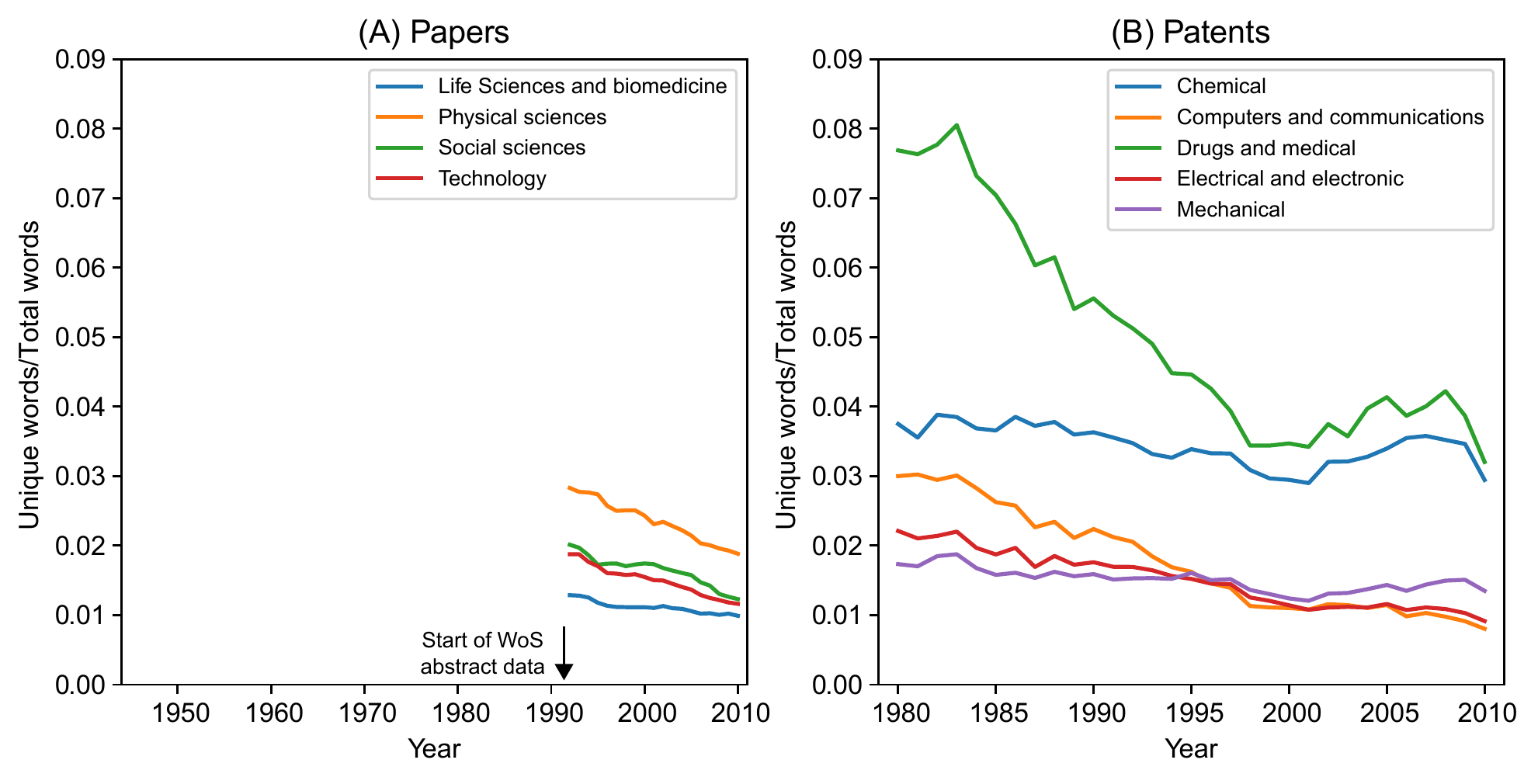}
\caption{\textbf{Diversity of language use in science and technology over time.} This figure shows changes in the ratio of unique to total words (also known as the type-token ratio) over time based on data from the abstracts of papers (\textbf{A}) and patents (\textbf{B}). For papers, lines correspond to WoS research areas; for patents, lines correspond to NBER technology categories. For paper abstracts, lines begin in 1992 because WoS does not reliably record abstracts for papers published prior to the early 1990s. The ratio of unique to total words is computed separately by field (i.e., the uniqueness of words and total word counts are determined within WoS research areas and NBER technology categories). If disruption is decreasing, we may plausibly expect to see a decrease in the diversity of words used by scientists and inventors, as discoveries and inventions will be less likely to create departures from the status quo, and will therefore be less likely to need to introduce new terminology. For both papers and patents, we observe declining diversity in word use over time, which is consistent with this expectation and corroborates our findings using the CD index.}
\label{figure:LexicalDiversityAbs}
\end{figure}
\pagebreak

\begin{figure}[H]
\includegraphics[width=\textwidth]{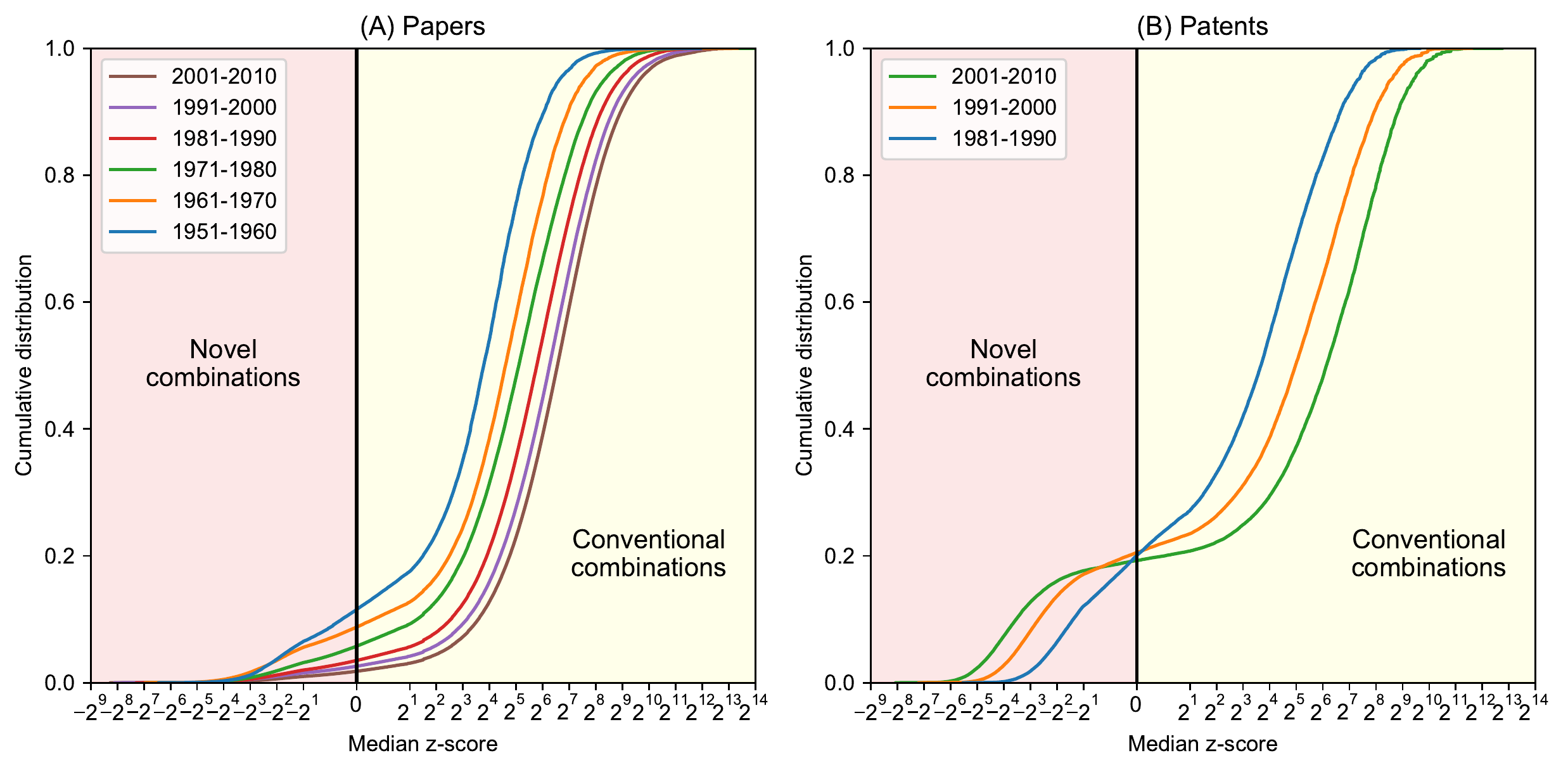}
\caption{\textbf{Declining combinatorial novelty.} This figure shows changing patterns in the combinatorial novelty/conventionality of papers (\textbf{A}) and patents (\textbf{B}), using the measure of ``atypical combinations'' proposed in \autocite{uzzi2013atypical}. The measure quantifies the degree to which the prior work cited by a paper or patent would be expected by chance. For papers, we follow \autocite{uzzi2013atypical} and consider combinations of cited journals. For example, if a paper made three citations to prior work, and that work was published in three different journals---\emph{Nature}, \emph{PNAS}, and \emph{Science}---then there are three combinations---\emph{Nature} $\times$ \emph{PNAS}, \emph{Nature} $\times$ \emph{Science}, and \emph{Science} $\times$ \emph{PNAS}. To determine the degree to which each combination would be expected by chance, the frequency of observed pairings are compared to those in 10 ``rewired'' copies of the overall citation network, using a z-score. For patents, there is no natural analogue to journals, and therefore we consider pairings of primary United States Patent Classification (USPC) system codes. We present the results of this analysis following the approach of \autocite{uzzi2013atypical}, who plotted the cumulative distribution function of their novelty measure, with separate lines for groups of interest (which in our case correspond to the decade of publication). In general, there is a rightward shift in the cumulative distributions over time, suggesting that for both papers and patents, combinations are more conventional than would be expected by chance, consistent with what we would anticipate based on our results using the CD index. For patents, there is also a smaller shift in the opposite direction on the left side of the distribution, suggesting that novel patents in recent decades are somewhat more novel than novel patents in earlier decades. Overall, however, the bulk of the distribution is moving to the right, indicating greater conventionality.}
\label{figure:AtypicalCombinations}
\end{figure}
\pagebreak

\begin{figure}[H]
\includegraphics[width=\textwidth]{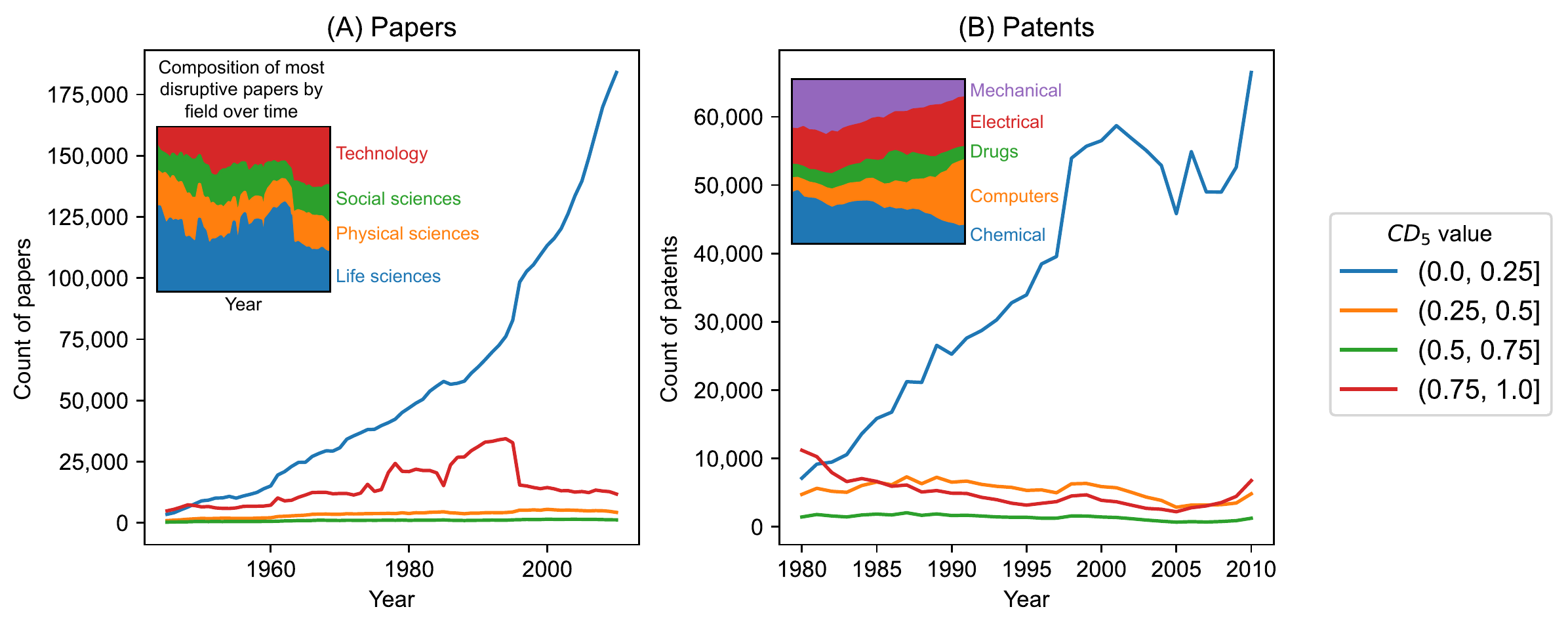}
\caption{\textbf{Persistence of major breakthroughs.} This figure shows the number of disruptive papers (\textbf{A}) and patents (\textbf{B}) across four different ranges of the $CD_5$ index (papers and patents with $CD_5$ values in the range [-1.0, 0.0) are not represented in the figure). Lines correspond to different levels of disruptiveness measured by the $CD_5$ index. Despite substantial increases in the numbers of papers and patents published each year, there is little change in the number of highly disruptive papers and patents, as evidenced by the relatively flat red, green, and orange lines. This pattern helps to account for simultaneous observations of both aggregate evidence of slowing innovative activity and seemingly major breakthroughs in many fields of science and technology. The inset plots show the composition of the most disruptive papers and patents (defined as those with $CD_5$ values $>0.25$) by field over time. The observed stability in the absolute number of highly disruptive papers and patents holds despite considerable churn in the underlying fields of science and technology responsible for producing those works.}
\label{figure:CDoutliers}
\end{figure}
\pagebreak

\begin{figure}[H]
\includegraphics[width=0.75\textwidth]{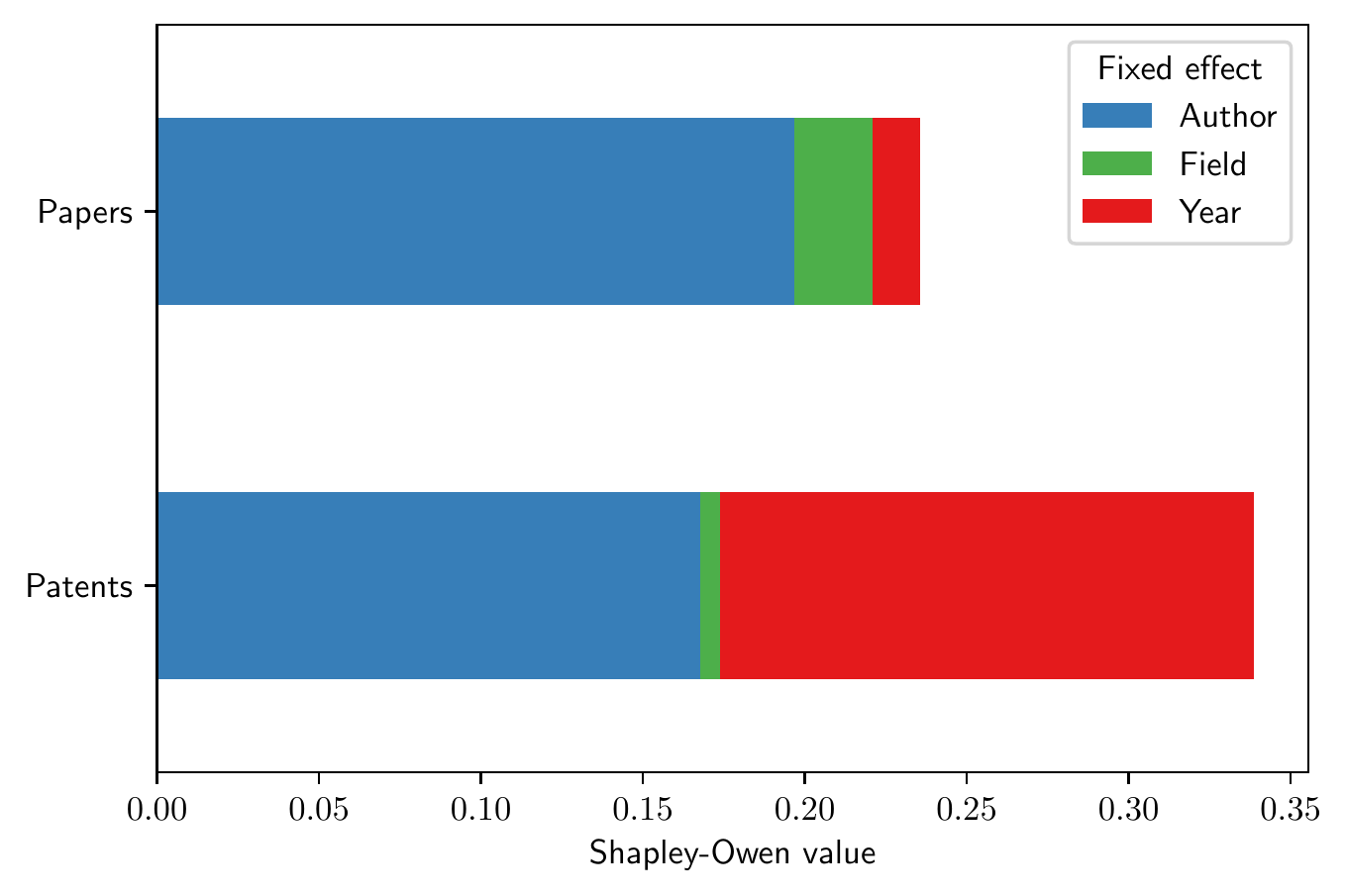}
\caption{\textbf{Contribution of field, year, and author effects.} This figure shows the relative contribution of field, year, and author fixed effects to the adjusted $R^2$ in regression models predicting the $CD_5$ index. The top bar shows the results for papers; the bottom bar shows the results for patents. The results suggest that for both papers and patents, stable characteristics of authors contribute significantly to patterns of disruptiveness. Moreover, relatively little of the variation is accounted for by field-specific factors}
\label{figure:r2contribution}
\end{figure}
\pagebreak

\begin{figure}[H]
\includegraphics[width=\textwidth]{raw/plot_nobel_nsp_combined.pdf}
\caption{\textbf{CD index of high quality science over time.} This figure shows changes in $CD_5$ over time for papers published in \textit{Nature, PNAS,} and \textit{Science} (\textbf{A}) and Nobel-Prize-winning papers (\textbf{B}). Colors indicate the three different journals in \textbf{A}; colors indicate the three different fields that receive the Nobel Prize in \textbf{B}. For historical completeness, we plot the $CD_5$ index scores for all Nobel papers back to 1900 (the first year in which the prize was awarded); however, our main analyses begin in the post-1945 era, when the Web of Science data are generally more reliable. The figure indicates that changes in the quality of published science over time is unlikely to be responsible for the decline in disruption.}
\label{figure:CDnobel}
\end{figure}
\pagebreak

\begin{figure}[H]
\includegraphics[width=0.75\textwidth]{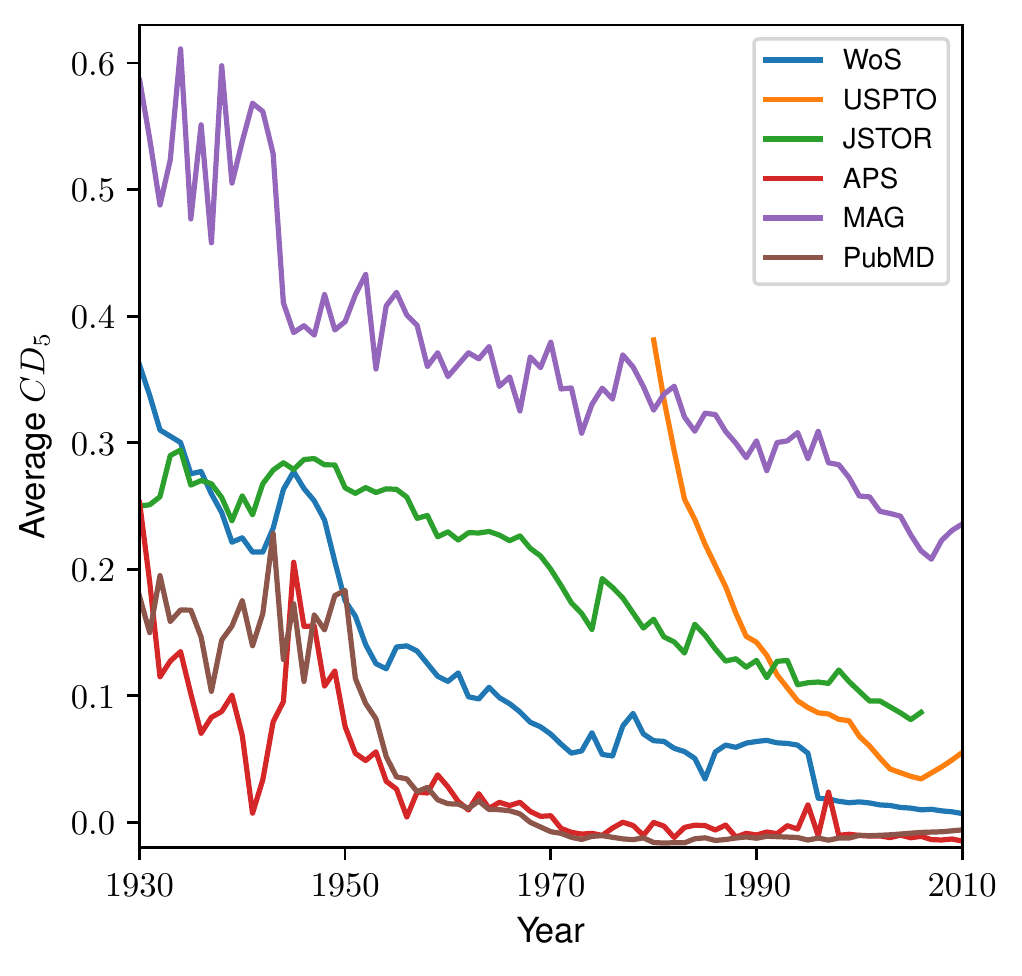}
\caption{\textbf{CD index over time across other data sources.} This figure shows changes in $CD_5$ over time across four additional data sources (the WoS and USPTO lines are included for reference): JSTOR, the American Physical Society corpus, Microsoft Academic Graph, and PubMD. Colors indicate the six different data sources. The figure indicates that the decline in disruption is unlikely to be driven by our sample choice of WoS papers and USPTO patents.}
\label{figure:CD_othersample}
\end{figure}
\pagebreak

\begin{figure}[H]
\includegraphics[width=\textwidth]{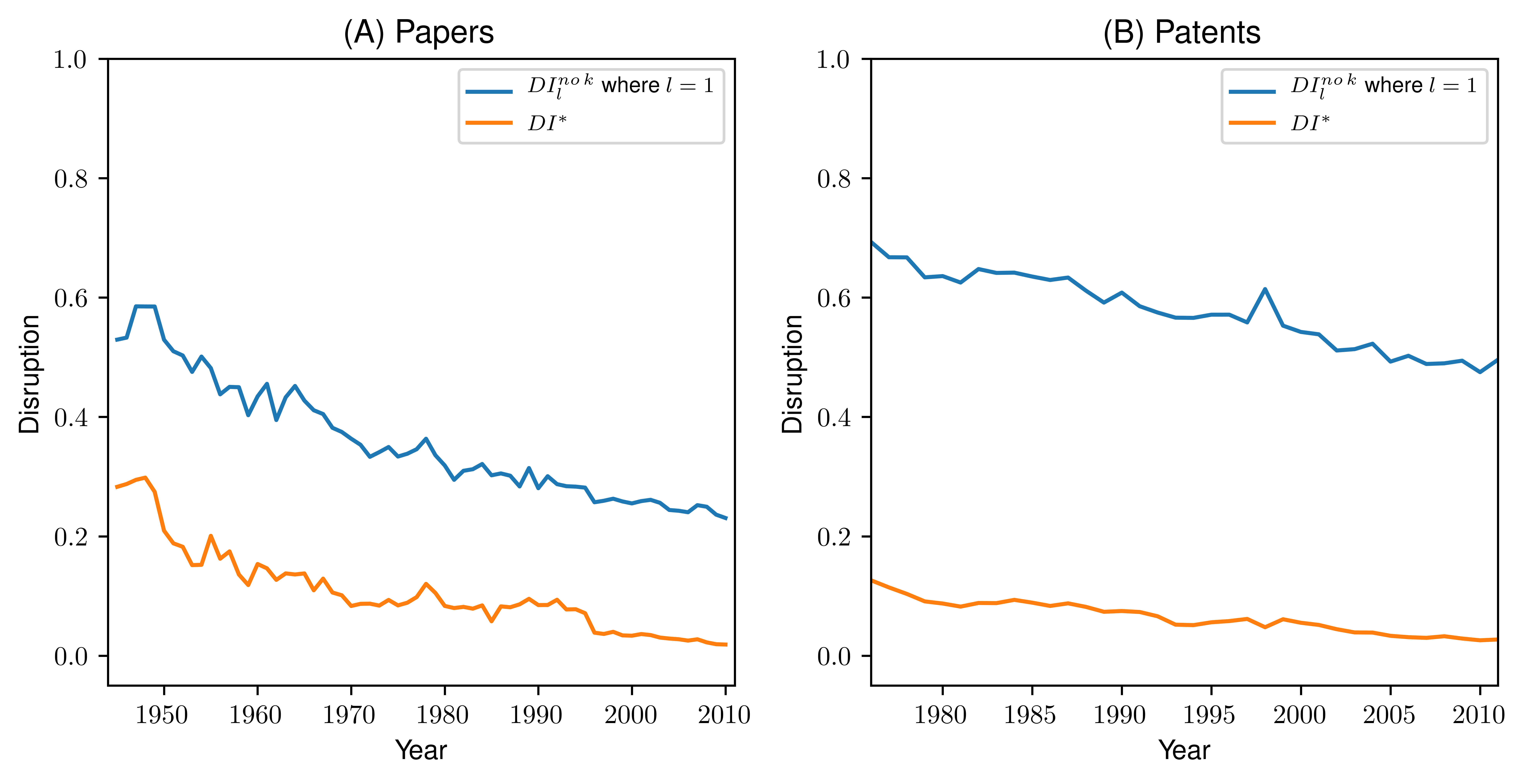}
\caption{\textbf{Alternative measures of disruption} This figure shows the decline in the disruption of papers (\textbf{A}) and patents (\textbf{B}) based on two alternative measures of disruption. The blue lines calculate disruption using a measure proposed in Bornmann et al.\autocite{bornmann2020disruption}, $DI_{l}^{no\,k}$ where $l = 1$, which makes the measure more resilient to marginal changes in the number of papers or patents that only cite the focal work's references. The orange lines calculate disruption using a measure proposed in Leydesdorff et al.\autocite{leydesdorff2021proposal}, $DI^*$, which makes the measure less sensitive to small changes in the forward citation patterns of papers or patents that make no backward citations. With both alternative measures, we observe decreases in disruption for papers and patents, suggesting that the decline is not an artifact of the $CD_5$ measure.}
\label{figure:altD}
\end{figure}
\pagebreak

\begin{figure}[H]
\includegraphics[width=\textwidth]{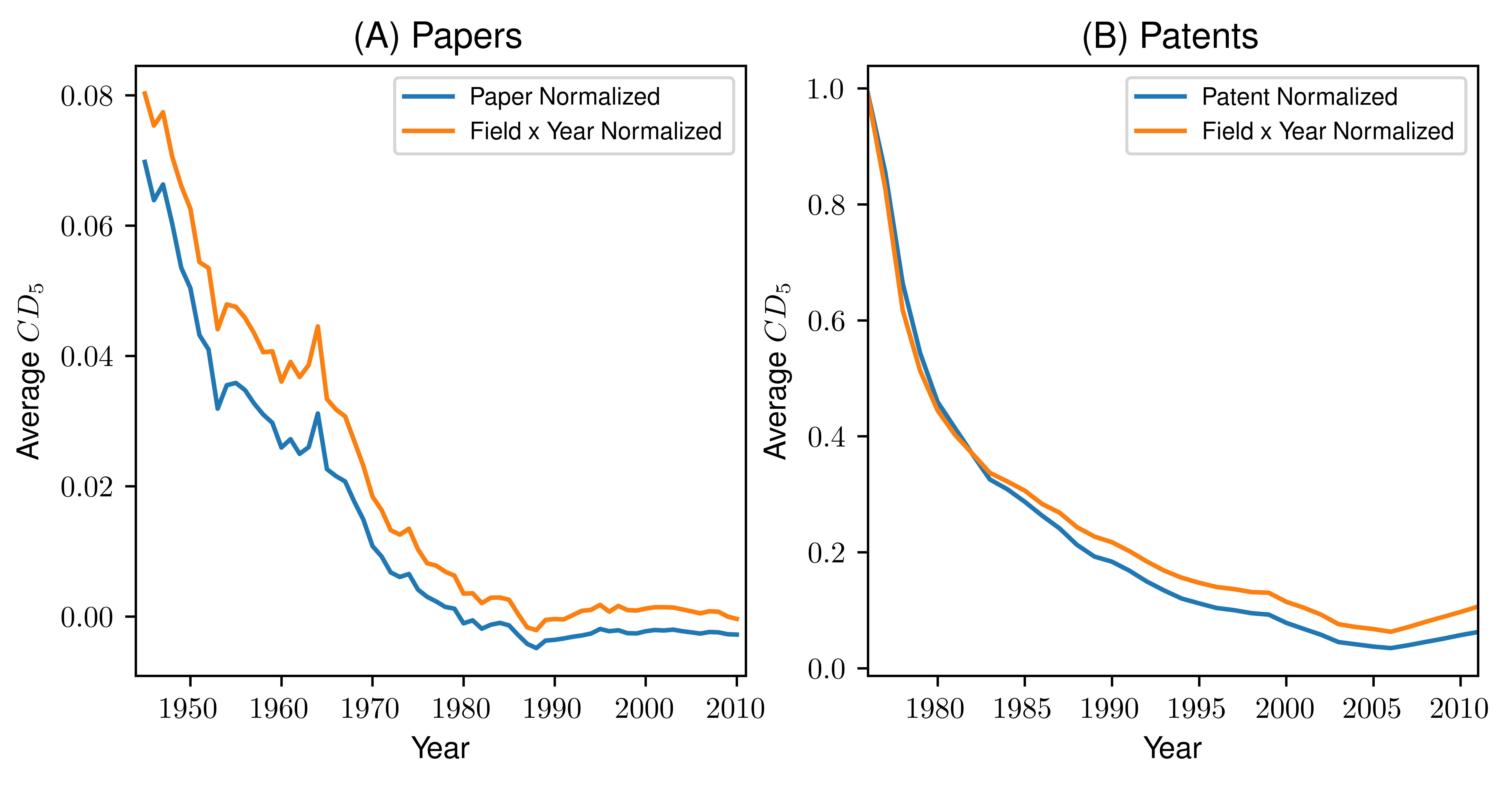}
\caption{\textbf{Paper and Field $\times$ year normalized versions of the $CD_5$} This figure shows the decline in the disruption of papers (\textbf{A}) and patents (\textbf{B}) based on two normalized versions of the $CD_5$ over time. Both normalization approaches are aimed at addressing the potential for $CD_5$ to deflate over time due to the tendency for more recent work to cite greater amounts of prior work. The blue lines indicate the value of $CD_5$ normalized at the paper level (``Paper normalized''), which takes into account the number of citations made by the focal paper or patent. The orange lines indicate the value of $CD_5$ normalized at the field and year level (``Field $\times$ year normalized''), which takes into account the mean number of citations made by the a paper or patent in the focal field and year. Even after normalizing the CD index at multiple levels, we continue to observe a decline in the disruption of both papers and patents across time suggesting that other temporal trends are unlikely to be responsible for the decreasing disruption.}
\label{figure:cd_norm}
\end{figure}
\pagebreak

\begin{figure}[H]
\includegraphics[width=6in]{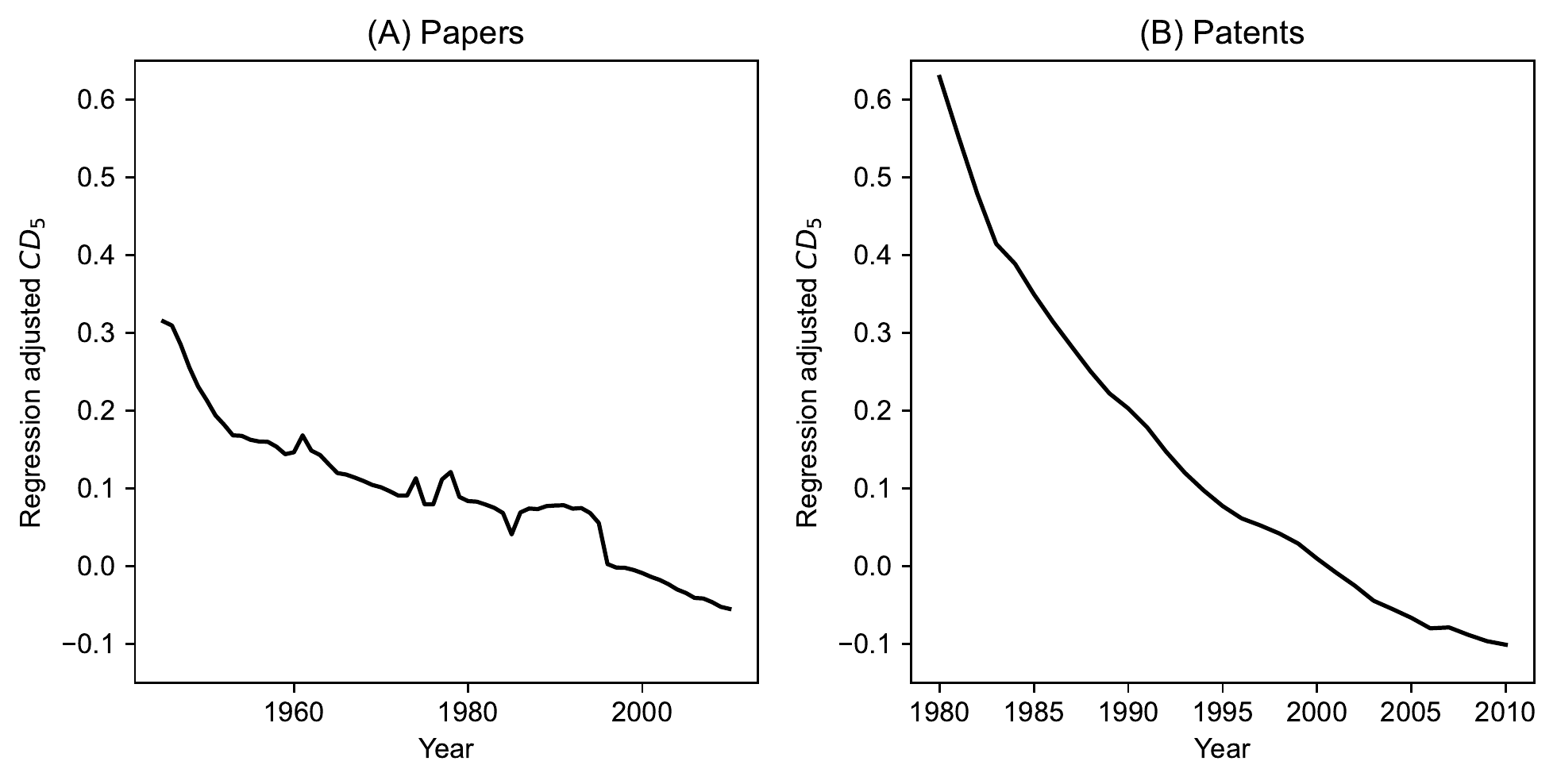}
\caption{\textbf{Predicted values of the $CD_5$ index by year after adjusting for patterns in publication, citation, and authorship practices.} This figure presents predicted values of the $CD_5$ index based on regressions reported in Models 4 (papers) and 8 (patents) of Supplementary Table 2. These models include adjustments for several field $\times$ year level---\emph{Number of new papers/patents}, \emph{Mean number of papers/patents cited}, \emph{Mean number of authors/inventors per paper/patent}---and paper/patent-level---\emph{Number of papers/patents cited}---characteristics, thereby enabling more robust comparisons in patterns of disruptive science and technology over the long time period spanned by our study. Predictions are made separately for each of the year indicators included in the models; we then connect these separate predictions with lines to facilitate interpretation and to enable comparisons with the raw values of the $CD_5$ index shown in the main text.} 
\label{figure:CDadjusted}
\end{figure}

\begin{figure}[H]
\includegraphics[width=\textwidth]{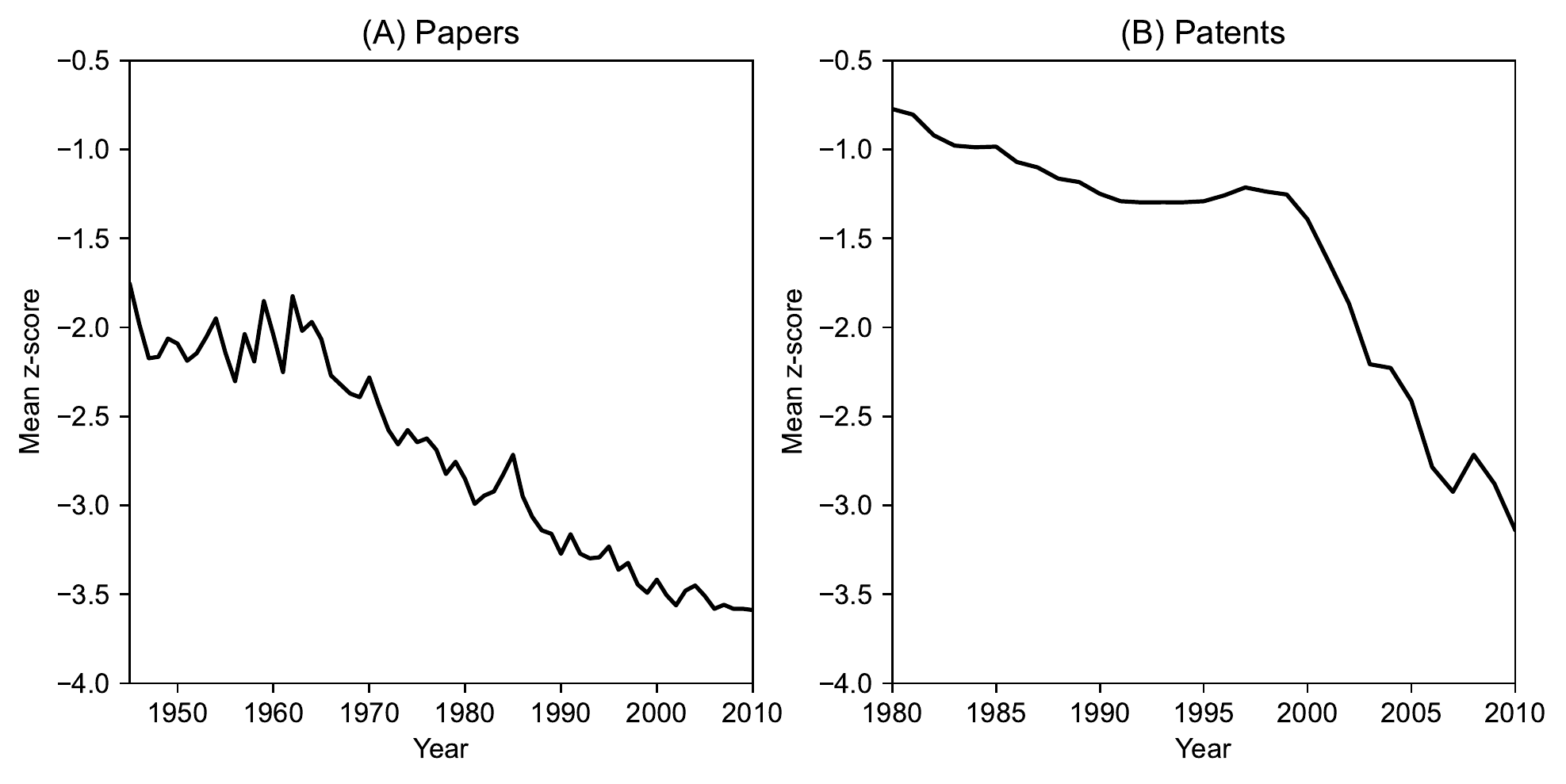}
\caption{\textbf{Comparison of the observed values of the $CD_5$ index to those obtained from randomly rewired citation networks.} This figure plots z-scores that compare the values of the $CD_5$ index obtained from the observed citation networks to those obtained from randomly rewired copies of the observed networks. The motivation for this analysis is to determine whether the observed declines in the $CD_5$ index could result more from structural changes in the citation networks of science and technology (e.g., increases in the number of papers/patents, increases in the number of citations made by papers/patents) than from a social process of disruption. For both papers (\textbf{A}) and patents (\textbf{B}), we find that the observed values of the $CD_5$ index tend to be lower than expected by chance (based on comparisons with the rewired networks), and moreover, the gap between the observed $CD_5$ index values and those from the randomly rewired networks is increasing over time, consistent with our findings of a decline in disruptive science and technology.}
\label{figure:CDrewire}
\end{figure}
\pagebreak

\begin{figure}[H]
\includegraphics[width=\textwidth]{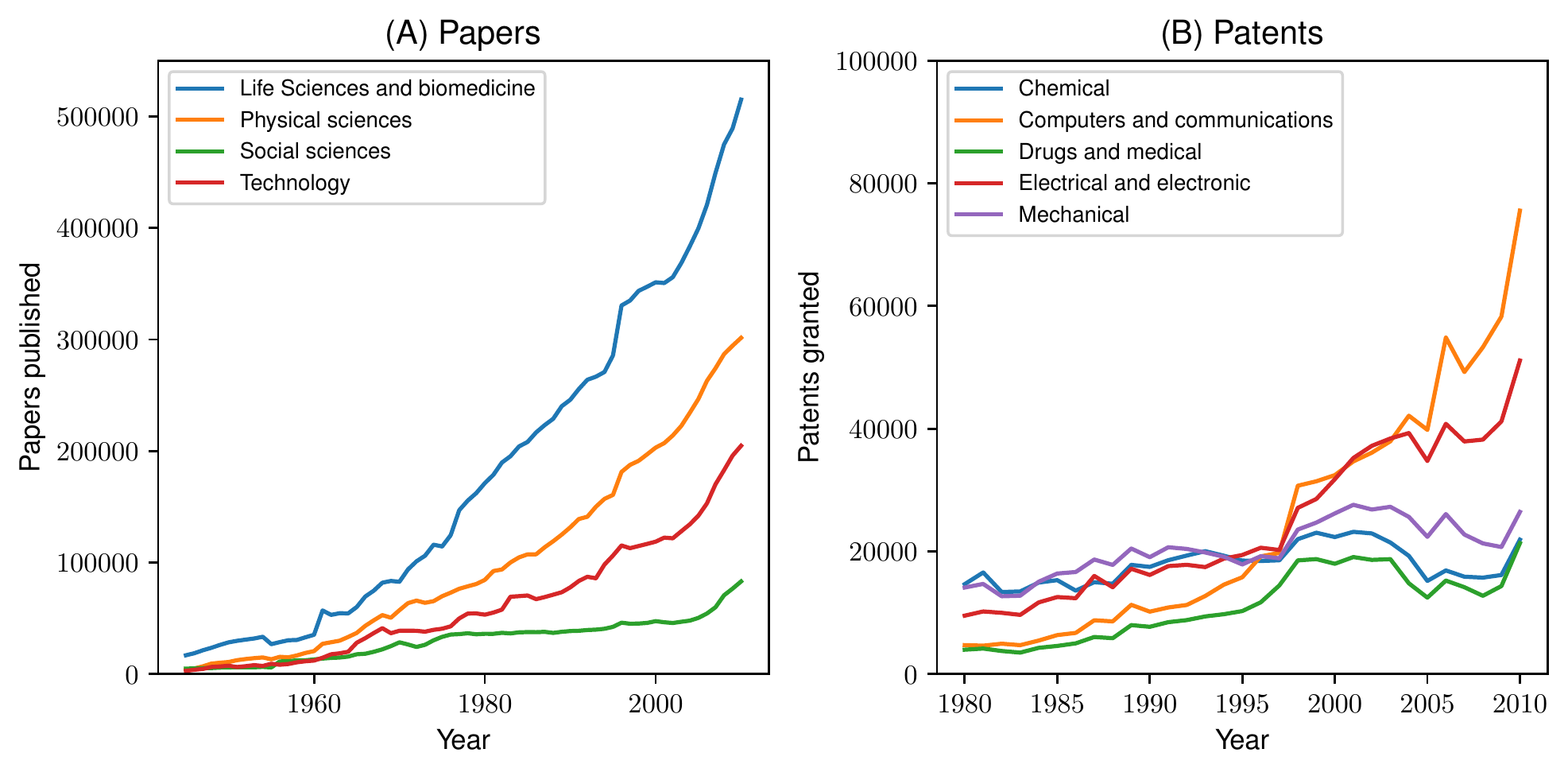}
\caption{\textbf{Growth of scientific and technological knowledge.} This figure shows the number of papers published (\textbf{A}) and patents granted (\textbf{B}) over time. For papers, lines correspond to WoS research areas; for patents, lines correspond to NBER technology categories.}
\label{figure:CDvolume}
\end{figure}
\pagebreak

\newgeometry{top=0.5in, bottom=0.5in, left=0.5in, right=0.5in}

{

\thispagestyle{empty}

\setlength{\tabcolsep}{3pt}


\begin{table}[htbp]\centering
\footnotesize
\begin{threeparttable}
\def\sym#1{\ifmmode^{#1}\else\(^{#1}\)\fi}
\caption{Summary statistics for variables used in regression analyses \\ WoS/Papers}
\label{table:SummaryStatisticsPapers}
\begin{tabular}{l*{7}{c}}
\toprule

Variable	&	$N$	&	$N_{unique}$	&	Mean & SD	&	Min	&	Max	& Level of measurement \\ \midrule

 Use of existing knowledge & \\   
 Diversity of work cited &  24578889 & 8820 & 0.98 & 0.01 & 0.88 & 1.00 & Field $\times$ year \\  
 Ratio of self-citations to total work cited &  21722268 & 5767 & 0.16 & 0.20 & 0.00 & 1.00 & Publication \\  
 Mean age of work cited &  21722268 & 74586 & 9.22 & 5.55 & 0.00 & 110.00 & Publication \\  
 Dispersion in age of work cited &  21722268 & 2610114 & 6.32 & 4.22 & 0.00 & 53.00 & Publication \\  
 \midrule   
 Controls & \\   
 Mean age of team members &  24585080 & 22615 & 13.67 & 10.75 & 0.00 & 80.00 &  Publication \\  
 Mean number of prior works produced by team members &  24585080 & 58738 & 35.43 & 55.91 & 0.00 & 2381.00 & Publication  \\  
 \midrule   
 Outcomes & \\   
 CD$\textsubscript{5}$ index &  22432124 & 848544 & 0.04 & 0.22 & -1.00 & 1.00 & Publication \\  
 \midrule   
 Fixed effects & \\   
 Publication year &  24585080 & 66 & -- & -- & -- & -- & Year \\  
 WoS extended subject &  24585080 & 150 & -- & -- & -- & -- & Field \\

\bottomrule
\end{tabular}

\end{threeparttable}
\end{table}

}%




{

\thispagestyle{empty}

\setlength{\tabcolsep}{3pt}


\begin{table}[htbp]\centering
\footnotesize
\begin{threeparttable}
\def\sym#1{\ifmmode^{#1}\else\(^{#1}\)\fi}
\caption*{USPTO/Patents}
\label{table:SummaryStatisticsPatents}
\begin{tabular}{l*{7}{c}}
\toprule

Variable	&	$N$	&	$N_{unique}$	&	Mean & SD	&	Min	&	Max	& Level of measurement \\ \midrule

 Use of existing knowledge & \\   
 Diversity of work cited &  3911466 & 1278 & 0.98 & 0.01 & 0.84 & 1.00 & Field $\times$ year \\  
 Ratio of self-citations to total work cited &  3485975 & 6121 & 0.07 & 0.18 & 0.00 & 1.00 & Publication \\  
 Mean age of work cited &  3485975 & 47944 & 7.67 & 4.15 & -36.00 & 34.00 & Publication \\  
 Dispersion in age of work cited &  3485975 & 282903 & 3.35 & 2.49 & 0.00 & 19.00 & Publication \\  
 \midrule   
 Controls & \\   
 Mean age of team members &  3911536 & 2115 & 5.60 & 5.93 & 0.00 & 34.00 & Publication \\  
 Mean number of prior works produced by team members &  3911536 & 8339 & 10.51 & 49.70 & 0.00 & 2852.00 & Publication \\  
 \midrule   
 Outcomes & \\   
 CD$\textsubscript{5}$ index &  3681322 & 58721 & 0.12 & 0.30 & -1.00 & 1.00 & Publication \\  
 \midrule   
 Fixed effects & \\   
 Publication year &  3911536 & 35 & -- & -- & -- & -- & Year \\  
 NBER technology subcategory &  3911536 & 38 & -- & -- & -- & -- & Field \\

\bottomrule
\end{tabular}

\end{threeparttable}
\end{table}

}%


\pagebreak


\begin{landscape}

\thispagestyle{empty}
\pagestyle{empty}

{

\scriptsize

\setlength{\tabcolsep}{2pt}

\renewcommand{\arraystretch}{0.85}

\begin{table}[htbp]\centering
\begin{adjustbox}{max width=\textwidth}
\begin{threeparttable}
\def\sym#1{\ifmmode^{#1}\else\(^{#1}\)\fi}
\caption{Regression models of disruptiveness and the use of prior knowledge}
\label{table:MAIN}
\begin{tabular}{l*{12}{c}}
\toprule

 
                                             &\multicolumn{6}{c}{\shortstack{Sample: Patents View}}                                                            &\multicolumn{6}{c}{\shortstack{Sample: Web of Science}}                                                          \\\cmidrule(lr){2-7}\cmidrule(lr){8-13}
                                             &\multicolumn{1}{c}{(1)}   &\multicolumn{1}{c}{(2)}   &\multicolumn{1}{c}{(3)}   &\multicolumn{1}{c}{(4)}   &\multicolumn{1}{c}{(5)}   &\multicolumn{1}{c}{(6)}   &\multicolumn{1}{c}{(7)}   &\multicolumn{1}{c}{(8)}   &\multicolumn{1}{c}{(9)}   &\multicolumn{1}{c}{(10)}   &\multicolumn{1}{c}{(11)}   &\multicolumn{1}{c}{(12)}   \\
\midrule
Diversity of work cited                      &         0.1151***&         0.1119***&         0.0873***&         1.3737***&         0.0688***&         0.0692***&         0.3293***&         0.3339***&         0.1574***&         0.4578***&         0.1583***&         0.1587***\\
                                             &       (0.0158)   &       (0.0158)   &       (0.0156)   &       (0.0080)   &       (0.0156)   &       (0.0156)   &       (0.0062)   &       (0.0062)   &       (0.0061)   &       (0.0025)   &       (0.0061)   &       (0.0061)   \\
Ratio of self-citations to total work cited  &                  &        -0.0606***&        -0.0557***&        -0.0671***&        -0.0585***&        -0.0597***&                  &        -0.0191***&        -0.0091***&        -0.0118***&        -0.0104***&        -0.0107***\\
                                             &                  &       (0.0009)   &       (0.0009)   &       (0.0009)   &       (0.0009)   &       (0.0009)   &                  &       (0.0003)   &       (0.0003)   &       (0.0003)   &       (0.0003)   &       (0.0003)   \\
Mean age of work cited                       &                  &                  &         0.0074***&         0.0008***&         0.0046***&         0.0046***&                  &                  &         0.0034***&         0.0027***&         0.0028***&         0.0028***\\
                                             &                  &                  &       (0.0000)   &       (0.0001)   &       (0.0001)   &       (0.0001)   &                  &                  &       (0.0000)   &       (0.0000)   &       (0.0000)   &       (0.0000)   \\
Dispersion in age of work cited              &                  &                  &        -0.0205***&        -0.0370***&        -0.0293***&        -0.0293***&                  &                  &        -0.0051***&        -0.0069***&        -0.0063***&        -0.0063***\\
                                             &                  &                  &       (0.0001)   &       (0.0001)   &       (0.0001)   &       (0.0001)   &                  &                  &       (0.0000)   &       (0.0000)   &       (0.0000)   &       (0.0000)   \\
Mean age of work cited $\times$ Dispersion in age of work cited&                  &                  &                  &         0.0013***&         0.0009***&         0.0009***&                  &                  &                  &         0.0001***&         0.0001***&         0.0001***\\
                                             &                  &                  &                  &       (0.0000)   &       (0.0000)   &       (0.0000)   &                  &                  &                  &       (0.0000)   &       (0.0000)   &       (0.0000)   \\
Mean age of team members                     &                  &                  &                  &                  &                  &         0.0002***&                  &                  &                  &                  &                  &         0.0000***\\
                                             &                  &                  &                  &                  &                  &       (0.0000)   &                  &                  &                  &                  &                  &       (0.0000)   \\
Mean number of prior works produced by team members&                  &                  &                  &                  &                  &        -0.0000***&                  &                  &                  &                  &                  &         0.0000***\\
                                             &                  &                  &                  &                  &                  &       (0.0000)   &                  &                  &                  &                  &                  &       (0.0000)   \\
\midrule Year fixed effects                  &            Yes   &            Yes   &            Yes   &             No   &            Yes   &            Yes   &            Yes   &            Yes   &            Yes   &             No   &            Yes   &            Yes   \\
Field fixed effects                          &            Yes   &            Yes   &            Yes   &             No   &            Yes   &            Yes   &            Yes   &            Yes   &            Yes   &             No   &            Yes   &            Yes   \\
\midrule
N                                            &        3433452   &        3433452   &        3433452   &        3433452   &        3433452   &        3433452   &       21553305   &       21553305   &       21553305   &       21553305   &       21553305   &       21553305   \\
R2                                           &           0.06   &           0.06   &           0.10   &           0.08   &           0.10   &           0.10   &           0.02   &           0.02   &           0.04   &           0.03   &           0.04   &           0.04   \\

\bottomrule

\end{tabular}

\begin{tablenotes}
\item \emph{Notes:} This table evaluates the relationship between different measures of the use of prior scientific and technological knowledge and the $CD_5$ index. Estimates are from ordinary-least-squares regressions. Robust standard errors are shown in parentheses; $p$-values correspond to two-tailed tests. 
\item {*}p$<$0.1 {**}p$<$0.05; {***}p$<$0.01
\end{tablenotes}

\end{threeparttable}
\end{adjustbox}
\end{table}

}%

\end{landscape}

\restoregeometry

\clearpage 
\section*{Supplementary Information}
\begin{refsection}
\renewcommand{\thesubsection}{\arabic{subsection}.}
\setcounter{subsection}{0}
\setcounter{table}{0} 
\setcounter{figure}{0} 
\renewcommand{\figurename}{Supplementary Figure}
\renewcommand{\tablename}{Supplementary Table}

\subsection{Changes in the diversity and novelty of word use over time} We examined changes in the diversity of and novelty of word use in papers and patents over time. Our rationale for these analyses is that decreases in disruption should be associated with decreases in the diversity of words used in science and technology. Disruptive discoveries and inventions create departures from the status quo, rendering their predecessors less useful. While this pattern alone may have the effect of reducing the diversity of words used, disruptive discoveries and inventions are also likely to introduce new words or combinations of words; part of the way that disruptive discoveries and inventions render their predecessors less useful is by introducing ideas that are more useful than those that came before, which are likely to require new words or combinations of existing words to describe. Taken together with the long memory of language (i.e., even obsolete words are still occasionally used\autocite{michel2011quantitative})\footnote{For example, in the WoS data, we find that words like ``phlogiston'', ``aether'', ``preformism'', and ``lamarkian'' continued to appear in the titles of scientific papers (though in much lower frequencies) well after their associated scientific theories were superseded.) We identified these terms from the Wikipedia page on “Superseded theories in science”: \url{https://en.wikipedia.org/wiki/Superseded_theories_in_science}.}, we therefore anticipate a positive association between disruption and the diversity and combinatorial novelty of word use by scientists and inventors. Thus, to the extent that our observations of decreasing disruption hold, we should see a decline in the diversity of words and novelty of their pairings over time.

To evaluate for such changes, we pulled all titles and abstracts for papers and patents in our sample from Web of Science and Patents View. For titles, there was very little missing data in either Web of Science or Patents View, with titles absent in fewer than 0.01\% of cases in both the former and the latter. For abstracts, Patents View also provides highly complete coverage, with only 0.32\%  of cases missing. Web of Science has less robust coverage of abstracts before the early 1990s; from 1945-1991, only 4.45\%  of papers in our sample include abstracts. Coverage is much better in later years; from 1992-2010, abstracts are included for 90.85\% of papers. We therefore limit our analyses of abstract data from WoS to the 1992-2010 period. 

After extracting paper and patent titles and abstracts, we completed a series of preprocessing steps using spaCy, an open-source, state-of-the-art Python package for natural language processing. To begin, we tokenized each title and abstract. From the resulting lists of tokens, we then excluded those that were tagged by spaCy as stop words, tokens consisting only of digits or punctuation, and tokens that were shorter than three characters or longer than 250 characters in length. Next, we converted all remaining tokens to their lemmatized form and converted all letters to lowercase. Finally, we aggregated the resulting lists of tokens to the subfield $\times$ year level, separately for papers and patents and for titles and abstracts. 
 
We evaluate changes in the diversity of words used over time by computing, for each subfield $\times$ year, the \emph{type-token ratio}, a common measure of unique/total words. The type-token ratio is defined as the ratio of unique words to total words. We compute this measure separately for papers and patents and for titles and abstracts, at the level of the Web of Science research area (for papers) and NBER technology category (for patents). More specifically, for each field (i.e., research area or technology category) and each year, we divide the number of unique words appearing in titles by the total number of words appearing in titles (Figure 3A and 3D). We repeat this step for the number of unique and total words appearing in abstracts (Extended Data Figure 3). The measure attains its theoretical maximum when every word is used exactly once. Thus, higher values indicate greater diversity.\footnote{We find similar results when measuring unique/total words using normalized entropy.} 

We followed a similar approach to measure the novelty of word pairs. For each subfield $\times$ year, we identify all pairwise combinations of words appearing in the titles of papers and patents. We then remove from that set all pairwise combinations of words in the titles of papers and patents published in the focal subfield in all previous years, which yields our count of new combinations. We then obtain our final (Figure 3b and 3E) measure by dividing this count of new combinations by the total number of word combinations made in the titles of papers and patents in the focal subfield and focal year. Higher values indicate greater combinatorial novelty in word use. 

\subsection{Changes in word use over time} We also examined changes in the specific words used in papers and patents over time. Our rationale for these analyses is that the changes we observe in the CD index are likely to coincide with changes in approaches to discovery and invention, particularly the orientation of scientists and inventors towards prior knowledge. For example, to the extent that disruption is decreasing over time, it seems plausible that we will also observe decreases in words indicating the creation, discovery, or perception of new things. Similarly, it is also plausible that we will observe concomitant increases in the use of words that are more indicative of improvement, application, or assessment of existing things, which, consistent with the notion of consolidation, may reinforce existing streams of knowledge.

To evaluate changes in the use of specific words over time, we followed an approach similar to that described used in our analyses of unique/total words, using similar samples of papers and patents and preprocessing steps. To simplify the presentation, we limit our attention to words appearing in paper and patent titles, for which, as noted previously, we have more complete data. However, the patterns we report below are also observable in analyses using paper and patent abstracts. Prior work has studied word frequencies in paper and patent titles extensively, and they are generally thought to provide a good window into the nature of science and technology \autocite{milojevic2015quantifying}. For the present analyses, during preprocessing, we also assigned a part of speech tag to each lemma, after which we extracted all nouns, verbs, adjectives, and adverbs, which we anticipated would provide the most meaningful insights. At this stage, our data consisted of counts of lemmas by part of speech appearing in the titles of sample papers and patents. To facilitate analysis, we subsequently reshaped the data in a long-panel format, separately for papers and patents, where each row was uniquely identified by a document id $\times$ part of speech $\times$ token.

We then examined changes in the top 10 most frequently used words in paper and patent titles by decade. For patents, we present these word frequencies for the years 1980 and 2010; for papers, our time series is longer, and therefore we present frequencies for 1950 and 2010. Analyses (available upon request from the authors) that include additional years (e.g., for each decade) yielded consistent results.

To simplify the presentation and conserve space, we focus our reporting on the results for verbs, which also generally yielded more noteworthy patterns (the most frequent nouns were often topical in nature; the most frequent adverbs and adjectives tended to be general/stop words). 

 We find evidence of a qualitative shift in word use that is consistent with our quantitative findings on the decline in disruptive activity using the CD index. Figure 3 shows the most common verbs in paper (Figure 3C) and patent titles (Figure 3F) in the first and last decade of each sample. In earlier decades, for example, verbs evoking creation (e.g., ``produce'', ``form'', ``prepare'', ``make''), discovery (e.g., ``determine'', ``report''), and perception (e.g., ``measure'') are prevalent in both paper and patent titles. In later decades, these verbs are almost completely displaced by those more evocative of the improvement (e.g., ``improve'', ``enhance'', ``increase''), application (e.g., ``use'', ``include''), or assessment (e.g., ``associate'', ``mediate'', ``relate'') of existing scientific and technological knowledge and artifacts. Thus, we observe a decrease in the use of verbs indicative of processes of disruption and a simultaneous increase in the use of verbs indicative of processes of consolidation. 

These results are especially noteworthy when recalling that they are based on raw data, with no adjustment or transformation other than basic text preprocessing. Overall, then, the patterns offer compelling support for the findings we observe on the changing nature of science and technology using the CD index.

\pagebreak

\subsection{Disruptiveness and the growth of knowledge} 
Extended Data Figure 13 plots the number of new papers published and archived by WoS (Extended Data Figure 13A) and utility patents granted by the USPTO each year (Extended Data Figure 13B). Extended Data Figure 13 shows that there has been a sharp and consistent increase in the number of new papers and patents. The rate of new papers and patents added year by year seems to be accelerating, expanding the existing stock of knowledge rapidly, and thereby potentially placing a ``burden of knowledge'' on scientists and inventors.

Models in Supplementary Table 1 evaluate the relationship between the growth of knowledge and disruptiveness. Models 1 and 4 proxy for new knowledge based on the number of new works (papers or patents) produced in the focal year. Models 2 and 5 proxy for new knowledge based on the number of new works (papers or patents) produced in the five most recent years. Models 3 and 6 proxy for new knowledge based on the number of new works (papers or patents) produced in the ten most recent years. We find divergent results for papers and patents; for the former, there is a negative association between new knowledge and $CD_5$; for the latter, the association is positive. This divergent pattern motivates our subsequent analyses (Extended Data Table 2), on the use of prior knowledge and disruptiveness.

\pagebreak


\newgeometry{top=0.5in,bottom=0.5in, left=0.5in, right=0.5in}

\thispagestyle{empty}
\pagestyle{empty}

{

\setlength{\tabcolsep}{2pt}

\renewcommand{\arraystretch}{0.85}

\begin{table}[htbp]\centering
\begin{threeparttable}
\def\sym#1{\ifmmode^{#1}\else\(^{#1}\)\fi}
\caption{Regression models of disruptiveness and the growth of knowledge}
\label{table:BURDEN}
\footnotesize
\begin{tabular}{l*{8}{c}}
\toprule
                                             &\multicolumn{3}{c}{\shortstack{Sample: Patents View}}                                         &\multicolumn{3}{c}{\shortstack{Sample: Web of Science}}                                       \\\cmidrule(lr){2-4}\cmidrule(lr){5-7}
                                             &\multicolumn{1}{c}{(1)}   &\multicolumn{1}{c}{(2)}   &\multicolumn{1}{c}{(3)}   &\multicolumn{1}{c}{(4)}   &\multicolumn{1}{c}{(5)}   &\multicolumn{1}{c}{(6)}   \\
\midrule
Number of new works in the field during focal year (logged)    &         0.0036***&                  &                  &  -0.0042***      &                  &                  \\
                                                               &        (0.0004)  &                  &                  &        (0.0001)  &                  &                  \\
Number of new works in the field during past 5 years (logged)  &                  &         0.0026***&                  &                  &        -0.0019***&                  \\
                                                               &                  &         (0.0002) &                  &                  &          (0.0001)&                  \\
Number of new works in the field during past 10 years (logged) &                  &                  &         0.0033***&                  &                  &      -0.0019***  \\
                                                               &                  &                  &          (0.0002)&                  &                  &       (0.0001)   \\
Constant                                                       &        0.0291*** &       0.0064     &          -0.0095*&       0.0519***  &      0.0479**    &      0.0491***   \\
                                                               &       (0.0030)   &       (0.0044)   &       (0.0051)   &       (0.0016)   &       (0.0016)   &       (0.0017)   \\
\midrule Year fixed effects                                    &            Yes   &            Yes   &            Yes   &             Yes  &            Yes   &            Yes   \\
Field fixed effects                                            &            Yes   &            Yes   &            Yes   &             Yes  &            Yes   &            Yes   \\

\midrule
N                                                              &        3434055  &        3434055   &        3434055   &       2.16e+07   &        2.16e+07   &       2.16e+07   \\
R2                                                             &           0.06  &           0.06   &           0.06   &            0.02  &           0.02    &           0.02   \\\bottomrule
\end{tabular}
\begin{tablenotes}
\item \emph{Notes:} This table evaluates the relationship between the number of new works (papers or patents, a proxy for the growth of knowledge) and the $CD_5$ index. Estimates are from ordinary-least-squares regressions. Robust standard errors are shown in parentheses; $p$-values correspond to two-tailed tests. 
\item {*}p$<$0.1 {**}p$<$0.05; {***}p$<$0.01
\end{tablenotes}
\end{threeparttable}
\end{table}

}%


\pagebreak


\thispagestyle{empty}
\pagestyle{empty}

{

\setlength{\tabcolsep}{1.75pt}

\renewcommand{\arraystretch}{0.75}

\begin{table}[htbp]\centering
\begin{threeparttable}
\caption{Regression models of trends in the $CD_5$ index adjusted for publication, citation, and authorship practices}
\label{table:CDadjusted}
\scriptsize
\begin{tabular}{l cc @{\hspace{5\tabcolsep}} cc @{\hspace{5\tabcolsep}} cc @{\hspace{5\tabcolsep}} cc @{\hspace{5\tabcolsep}} cc @{\hspace{5\tabcolsep}} cc @{\hspace{5\tabcolsep}} cc @{\hspace{5\tabcolsep}} cc @{\hspace{5\tabcolsep}} cc }\toprule
 
                                             &\multicolumn{8}{c}{\shortstack{Sample: Web of Science}}                                                                                    &\multicolumn{8}{c}{\shortstack{Sample: Patents View}}                                                                                      \\\cmidrule(lr){2-9}\cmidrule(lr){10-17}
                                             &\multicolumn{2}{c}{\shortstack{(1)}}&\multicolumn{2}{c}{\shortstack{(2)}}&\multicolumn{2}{c}{\shortstack{(3)}}&\multicolumn{2}{c}{\shortstack{(4)}}&\multicolumn{2}{c}{\shortstack{(5)}}&\multicolumn{2}{c}{\shortstack{(6)}}&\multicolumn{2}{c}{\shortstack{(7)}}&\multicolumn{2}{c}{\shortstack{(8)}}\\\cmidrule{2-3}\cmidrule{4-5}\cmidrule{6-7}\cmidrule{8-9}\cmidrule{10-11}\cmidrule{12-13}\cmidrule{14-15}\cmidrule{16-17}
                                             &              b   &             se&              b   &             se&              b   &             se&              b   &             se&              b   &             se&              b   &             se&              b   &             se&              b   &             se\\
\midrule
Year=1946                                    &          -0.01***&           0.00&          -0.01***&           0.00&          -0.01***&           0.00&          -0.01***&           0.00&                  &               &                  &               &                  &               &                  &               \\
Year=1947                                    &          -0.02***&           0.00&          -0.02***&           0.00&          -0.02***&           0.00&          -0.02***&           0.00&                  &               &                  &               &                  &               &                  &               \\
Year=1948                                    &          -0.04***&           0.00&          -0.03***&           0.00&          -0.03***&           0.00&          -0.03***&           0.00&                  &               &                  &               &                  &               &                  &               \\
Year=1949                                    &          -0.07***&           0.00&          -0.07***&           0.00&          -0.07***&           0.00&          -0.07***&           0.00&                  &               &                  &               &                  &               &                  &               \\
Year=1950                                    &          -0.10***&           0.00&          -0.10***&           0.00&          -0.09***&           0.00&          -0.10***&           0.00&                  &               &                  &               &                  &               &                  &               \\
Year=1951                                    &          -0.11***&           0.00&          -0.11***&           0.00&          -0.11***&           0.00&          -0.11***&           0.00&                  &               &                  &               &                  &               &                  &               \\
Year=1952                                    &          -0.14***&           0.00&          -0.13***&           0.00&          -0.13***&           0.00&          -0.13***&           0.00&                  &               &                  &               &                  &               &                  &               \\
Year=1953                                    &          -0.15***&           0.00&          -0.15***&           0.00&          -0.14***&           0.00&          -0.15***&           0.00&                  &               &                  &               &                  &               &                  &               \\
Year=1954                                    &          -0.16***&           0.00&          -0.15***&           0.00&          -0.14***&           0.00&          -0.15***&           0.00&                  &               &                  &               &                  &               &                  &               \\
Year=1955                                    &          -0.14***&           0.00&          -0.13***&           0.00&          -0.13***&           0.00&          -0.14***&           0.00&                  &               &                  &               &                  &               &                  &               \\
Year=1956                                    &          -0.14***&           0.00&          -0.13***&           0.00&          -0.13***&           0.00&          -0.14***&           0.00&                  &               &                  &               &                  &               &                  &               \\
Year=1957                                    &          -0.14***&           0.00&          -0.14***&           0.00&          -0.13***&           0.00&          -0.14***&           0.00&                  &               &                  &               &                  &               &                  &               \\
Year=1958                                    &          -0.15***&           0.00&          -0.15***&           0.00&          -0.14***&           0.00&          -0.15***&           0.00&                  &               &                  &               &                  &               &                  &               \\
Year=1959                                    &          -0.16***&           0.00&          -0.16***&           0.00&          -0.15***&           0.00&          -0.16***&           0.00&                  &               &                  &               &                  &               &                  &               \\
Year=1960                                    &          -0.17***&           0.00&          -0.16***&           0.00&          -0.15***&           0.00&          -0.17***&           0.00&                  &               &                  &               &                  &               &                  &               \\
Year=1961                                    &          -0.16***&           0.00&          -0.15***&           0.00&          -0.15***&           0.00&          -0.16***&           0.00&                  &               &                  &               &                  &               &                  &               \\
Year=1962                                    &          -0.18***&           0.00&          -0.17***&           0.00&          -0.16***&           0.00&          -0.18***&           0.00&                  &               &                  &               &                  &               &                  &               \\
Year=1963                                    &          -0.18***&           0.00&          -0.17***&           0.00&          -0.16***&           0.00&          -0.18***&           0.00&                  &               &                  &               &                  &               &                  &               \\
Year=1964                                    &          -0.17***&           0.00&          -0.17***&           0.00&          -0.16***&           0.00&          -0.18***&           0.00&                  &               &                  &               &                  &               &                  &               \\
Year=1965                                    &          -0.18***&           0.00&          -0.18***&           0.00&          -0.16***&           0.00&          -0.19***&           0.00&                  &               &                  &               &                  &               &                  &               \\
Year=1966                                    &          -0.18***&           0.00&          -0.18***&           0.00&          -0.17***&           0.00&          -0.19***&           0.00&                  &               &                  &               &                  &               &                  &               \\
Year=1967                                    &          -0.19***&           0.00&          -0.19***&           0.00&          -0.17***&           0.00&          -0.20***&           0.00&                  &               &                  &               &                  &               &                  &               \\
Year=1968                                    &          -0.20***&           0.00&          -0.20***&           0.00&          -0.18***&           0.00&          -0.21***&           0.00&                  &               &                  &               &                  &               &                  &               \\
Year=1969                                    &          -0.20***&           0.00&          -0.20***&           0.00&          -0.19***&           0.00&          -0.22***&           0.00&                  &               &                  &               &                  &               &                  &               \\
Year=1970                                    &          -0.21***&           0.00&          -0.21***&           0.00&          -0.19***&           0.00&          -0.22***&           0.00&                  &               &                  &               &                  &               &                  &               \\
Year=1971                                    &          -0.22***&           0.00&          -0.21***&           0.00&          -0.20***&           0.00&          -0.23***&           0.00&                  &               &                  &               &                  &               &                  &               \\
Year=1972                                    &          -0.22***&           0.00&          -0.22***&           0.00&          -0.20***&           0.00&          -0.24***&           0.00&                  &               &                  &               &                  &               &                  &               \\
Year=1973                                    &          -0.22***&           0.00&          -0.22***&           0.00&          -0.20***&           0.00&          -0.24***&           0.00&                  &               &                  &               &                  &               &                  &               \\
Year=1974                                    &          -0.21***&           0.00&          -0.20***&           0.00&          -0.19***&           0.00&          -0.23***&           0.00&                  &               &                  &               &                  &               &                  &               \\
Year=1975                                    &          -0.22***&           0.00&          -0.22***&           0.00&          -0.20***&           0.00&          -0.25***&           0.00&                  &               &                  &               &                  &               &                  &               \\
Year=1976                                    &          -0.22***&           0.00&          -0.22***&           0.00&          -0.21***&           0.00&          -0.25***&           0.00&                  &               &                  &               &                  &               &                  &               \\
Year=1977                                    &          -0.20***&           0.00&          -0.20***&           0.00&          -0.18***&           0.00&          -0.23***&           0.00&                  &               &                  &               &                  &               &                  &               \\
Year=1978                                    &          -0.19***&           0.00&          -0.19***&           0.00&          -0.17***&           0.00&          -0.22***&           0.00&                  &               &                  &               &                  &               &                  &               \\
Year=1979                                    &          -0.21***&           0.00&          -0.21***&           0.00&          -0.19***&           0.00&          -0.24***&           0.00&                  &               &                  &               &                  &               &                  &               \\
Year=1980                                    &          -0.21***&           0.00&          -0.21***&           0.00&          -0.19***&           0.00&          -0.25***&           0.00&                  &               &                  &               &                  &               &                  &               \\
Year=1981                                    &          -0.21***&           0.00&          -0.21***&           0.00&          -0.19***&           0.00&          -0.25***&           0.00&          -0.05***&           0.00&          -0.05***&           0.00&          -0.05***&           0.00&          -0.05***&           0.00\\
Year=1982                                    &          -0.22***&           0.00&          -0.22***&           0.00&          -0.20***&           0.00&          -0.26***&           0.00&          -0.09***&           0.00&          -0.09***&           0.00&          -0.09***&           0.00&          -0.09***&           0.00\\
Year=1983                                    &          -0.22***&           0.00&          -0.22***&           0.00&          -0.20***&           0.00&          -0.27***&           0.00&          -0.13***&           0.00&          -0.13***&           0.00&          -0.12***&           0.00&          -0.13***&           0.00\\
Year=1984                                    &          -0.23***&           0.00&          -0.23***&           0.00&          -0.20***&           0.00&          -0.27***&           0.00&          -0.14***&           0.00&          -0.14***&           0.00&          -0.14***&           0.00&          -0.14***&           0.00\\
Year=1985                                    &          -0.24***&           0.00&          -0.24***&           0.00&          -0.22***&           0.00&          -0.29***&           0.00&          -0.16***&           0.00&          -0.16***&           0.00&          -0.16***&           0.00&          -0.16***&           0.00\\
Year=1986                                    &          -0.22***&           0.00&          -0.22***&           0.00&          -0.19***&           0.00&          -0.27***&           0.00&          -0.18***&           0.00&          -0.18***&           0.00&          -0.18***&           0.00&          -0.18***&           0.00\\
Year=1987                                    &          -0.22***&           0.00&          -0.22***&           0.00&          -0.19***&           0.00&          -0.27***&           0.00&          -0.19***&           0.00&          -0.19***&           0.00&          -0.19***&           0.00&          -0.20***&           0.00\\
Year=1988                                    &          -0.22***&           0.00&          -0.22***&           0.00&          -0.19***&           0.00&          -0.27***&           0.00&          -0.22***&           0.00&          -0.22***&           0.00&          -0.21***&           0.00&          -0.22***&           0.00\\
Year=1989                                    &          -0.21***&           0.00&          -0.21***&           0.00&          -0.19***&           0.00&          -0.27***&           0.00&          -0.23***&           0.00&          -0.23***&           0.00&          -0.23***&           0.00&          -0.24***&           0.00\\
Year=1990                                    &          -0.21***&           0.00&          -0.21***&           0.00&          -0.18***&           0.00&          -0.28***&           0.00&          -0.24***&           0.00&          -0.24***&           0.00&          -0.23***&           0.00&          -0.25***&           0.00\\
Year=1991                                    &          -0.21***&           0.00&          -0.21***&           0.00&          -0.18***&           0.00&          -0.28***&           0.00&          -0.25***&           0.00&          -0.25***&           0.00&          -0.25***&           0.00&          -0.26***&           0.00\\
Year=1992                                    &          -0.21***&           0.00&          -0.21***&           0.00&          -0.18***&           0.00&          -0.29***&           0.00&          -0.26***&           0.00&          -0.26***&           0.00&          -0.26***&           0.00&          -0.28***&           0.00\\
Year=1993                                    &          -0.21***&           0.00&          -0.21***&           0.00&          -0.18***&           0.00&          -0.29***&           0.00&          -0.27***&           0.00&          -0.27***&           0.00&          -0.27***&           0.00&          -0.29***&           0.00\\
Year=1994                                    &          -0.22***&           0.00&          -0.22***&           0.00&          -0.18***&           0.00&          -0.29***&           0.00&          -0.29***&           0.00&          -0.28***&           0.00&          -0.28***&           0.00&          -0.30***&           0.00\\
Year=1995                                    &          -0.22***&           0.00&          -0.22***&           0.00&          -0.19***&           0.00&          -0.31***&           0.00&          -0.29***&           0.00&          -0.29***&           0.00&          -0.28***&           0.00&          -0.30***&           0.00\\
Year=1996                                    &          -0.26***&           0.00&          -0.26***&           0.00&          -0.22***&           0.00&          -0.35***&           0.00&          -0.29***&           0.00&          -0.29***&           0.00&          -0.29***&           0.00&          -0.31***&           0.00\\
Year=1997                                    &          -0.26***&           0.00&          -0.26***&           0.00&          -0.22***&           0.00&          -0.35***&           0.00&          -0.30***&           0.00&          -0.30***&           0.00&          -0.29***&           0.00&          -0.31***&           0.00\\
Year=1998                                    &          -0.26***&           0.00&          -0.26***&           0.00&          -0.22***&           0.00&          -0.35***&           0.00&          -0.30***&           0.00&          -0.30***&           0.00&          -0.29***&           0.00&          -0.32***&           0.00\\
Year=1999                                    &          -0.26***&           0.00&          -0.26***&           0.00&          -0.22***&           0.00&          -0.36***&           0.00&          -0.30***&           0.00&          -0.30***&           0.00&          -0.29***&           0.00&          -0.33***&           0.00\\
Year=2000                                    &          -0.26***&           0.00&          -0.26***&           0.00&          -0.22***&           0.00&          -0.36***&           0.00&          -0.31***&           0.00&          -0.31***&           0.00&          -0.31***&           0.00&          -0.34***&           0.00\\
Year=2001                                    &          -0.26***&           0.00&          -0.26***&           0.00&          -0.22***&           0.00&          -0.36***&           0.00&          -0.32***&           0.00&          -0.32***&           0.00&          -0.31***&           0.00&          -0.35***&           0.00\\
Year=2002                                    &          -0.26***&           0.00&          -0.26***&           0.00&          -0.22***&           0.00&          -0.37***&           0.00&          -0.33***&           0.00&          -0.33***&           0.00&          -0.32***&           0.00&          -0.36***&           0.00\\
Year=2003                                    &          -0.26***&           0.00&          -0.26***&           0.00&          -0.22***&           0.00&          -0.37***&           0.00&          -0.34***&           0.00&          -0.34***&           0.00&          -0.33***&           0.00&          -0.37***&           0.00\\
Year=2004                                    &          -0.27***&           0.00&          -0.27***&           0.00&          -0.22***&           0.00&          -0.38***&           0.00&          -0.34***&           0.00&          -0.34***&           0.00&          -0.33***&           0.00&          -0.37***&           0.00\\
Year=2005                                    &          -0.27***&           0.00&          -0.27***&           0.00&          -0.22***&           0.00&          -0.39***&           0.00&          -0.34***&           0.00&          -0.35***&           0.00&          -0.33***&           0.00&          -0.37***&           0.00\\
Year=2006                                    &          -0.27***&           0.00&          -0.27***&           0.00&          -0.22***&           0.00&          -0.40***&           0.00&          -0.35***&           0.00&          -0.35***&           0.00&          -0.33***&           0.00&          -0.38***&           0.00\\
Year=2007                                    &          -0.27***&           0.00&          -0.27***&           0.00&          -0.22***&           0.00&          -0.40***&           0.00&          -0.34***&           0.00&          -0.34***&           0.00&          -0.33***&           0.00&          -0.38***&           0.00\\
Year=2008                                    &          -0.27***&           0.00&          -0.27***&           0.00&          -0.22***&           0.00&          -0.40***&           0.00&          -0.34***&           0.00&          -0.34***&           0.00&          -0.32***&           0.00&          -0.37***&           0.00\\
Year=2009                                    &          -0.27***&           0.00&          -0.28***&           0.00&          -0.22***&           0.00&          -0.41***&           0.00&          -0.33***&           0.00&          -0.33***&           0.00&          -0.32***&           0.00&          -0.37***&           0.00\\
Year=2010                                    &          -0.27***&           0.00&          -0.28***&           0.00&          -0.22***&           0.00&          -0.42***&           0.00&          -0.33***&           0.00&          -0.33***&           0.00&          -0.31***&           0.00&          -0.37***&           0.00\\
\\Constant                                   &           0.28***&           0.00&           0.28***&           0.00&           0.29***&           0.00&           0.22***&           0.00&           0.38***&           0.00&           0.38***&           0.00&           0.38***&           0.00&           0.32***&           0.00\\\midrule
Subfield fixed effects                       &             No   &               &            Yes   &               &            Yes   &               &            Yes   &               &             No   &               &            Yes   &               &            Yes   &               &            Yes   &               \\
Paper/patent-level controls                  &             No   &               &             No   &               &            Yes   &               &            Yes   &               &             No   &               &             No   &               &            Yes   &               &            Yes   &               \\
Field $\times$ year-level controls           &             No   &               &             No   &               &             No   &               &            Yes   &               &             No   &               &             No   &               &             No   &               &            Yes   &               \\
\midrule
N                                            &       22456096   &               &       22456096   &               &       22456096   &               &       22456096   &               &        2926923   &               &        2926923   &               &        2926923   &               &        2926923   &               \\
R2                                           &           0.03   &               &           0.07   &               &           0.10   &               &           0.11   &               &           0.08   &               &           0.09   &               &           0.09   &               &           0.10   &               \\
\midrule Wald tests of controls              &                  &               &                  &               &                  &               &                  &               &                  &               &                  &               &                  &               &                  &               \\
F                                            &                  &               &                  &               &       46534.52   &               &       36458.66   &               &                  &               &                  &               &        9198.56   &               &        2974.44   &               \\
d.f.                                         &                  &               &                  &               &           1.00   &               &           4.00   &               &                  &               &                  &               &           1.00   &               &           4.00   &               \\
p-value                                      &                  &               &                  &               &           0.00   &               &           0.00   &               &                  &               &                  &               &           0.00   &               &           0.00   &               \\
 
\bottomrule
\end{tabular} 
\begin{tablenotes}
\item \emph{Notes:} Estimates are from ordinary-least-squares regressions. Robust standard errors are shown in parentheses; p-values correspond to two-tailed tests. Models 1 and 5 are the baseline models, and include no adjustments. Models 2 and 6 adds fixed effects for subfield. Models 3-4 (papers) and 7-8 (patents) add the field $\times$ year- and paper/patent-level controls, including adjustments for field $\times$ year level---\emph{Number of new papers/patents}, \emph{Mean number of papers/patents cited}, \emph{Mean number of authors/inventors per paper/patent}---and paper/patent-level---\emph{Number of papers/patents cited}---characteristics. The reference categories for the year indicators are 1945 and 1980 for papers and patents, respectively. The Wald tests, shown below the models including the control variables, test the null hypothesis that parameters for the control variables are simultaneously zero. Rejection of the null hypothesis is evidence that the control variables add to the explanatory power of the model.
\item {*}p$<$0.1 {**}p$<$0.05; {***}p$<$0.01
\end{tablenotes}
\end{threeparttable}
\end{table}

}%


\restoregeometry

\pagebreak

\printbibliography[heading=subbibliography]
\end{refsection}

\end{document}